\documentclass[useAMS,usenatbib,twoside,a4paper,iop,numberedappendix,appendixfloats]{emulateapj}

\usepackage{subfloat} 
\usepackage{epsfig}
\usepackage{amsmath}
\usepackage{natbib}
\usepackage{graphicx}
\usepackage{color}
\usepackage{rotating} 
\newcommand{\Msol}{\mbox{$M_{\odot}$}}
\newcommand{\grey}[1]{\textcolor[gray]{0.45}{#1}}
\usepackage{epstopdf}

\def\lta{{\>\rlap{\raise2pt\hbox{$<$}}\lower3pt\hbox{$\sim$}\>}}
\def\gta{{\>\rlap{\raise2pt\hbox{$>$}}\lower3pt\hbox{$\sim$}\>}}

\def\gsim{\mathrel{\raise0.35ex\hbox{$\scriptstyle >$}\kern-0.6em
\lower0.40ex\hbox{{$\scriptstyle \sim$}}}}
\def\lsim{\mathrel{\raise0.35ex\hbox{$\scriptstyle <$}\kern-0.6em
\lower0.40ex\hbox{{$\scriptstyle \sim$}}}}

\def\som{\mbox{M}_{\odot}}

\begin{document}

\title[]{ZENS IV. Similar Morphological Changes associated with \\ Mass- and Environment-Quenching, and the Relative importance of \\ Bulge Growth versus the Fading of Disks}

\author{C. M. Carollo\altaffilmark{1,$\star$}, 
A. Cibinel\altaffilmark{1,$\dagger$},
S. J. Lilly\altaffilmark{1}, 
A. Pipino\altaffilmark{1}, 
S. Bonoli\altaffilmark{2},
A. Finoguenov\altaffilmark{3},
F. Miniati\altaffilmark{1},
P. Norberg\altaffilmark{4},
J. D. Silverman\altaffilmark{5}
}
 
\altaffiltext{$\star$}{E-mail: \texttt{marcella@phys.ethz.ch}}
\altaffiltext{$\dagger$}{Current address:  CEA Saclay, DSM/Irfu/Service d'Astrophysique, Orme des Merisiers, F-91191 Gif-sur-Yvette Cedex, France}
\altaffiltext{1}{Institute for Astronomy, ETH Zurich, CH-8093 Zurich, Switzerland}
\altaffiltext{2}{Centro de Estudios de Fisica del Cosmos de Aragon, Spain}
\altaffiltext{3}{Department of Physics, University of Helsinki, FI-00014 Helsinki, Finland}

\altaffiltext{4}{Institute for Computational Cosmology, Department of
Physics, Durham University, South Road, Durham DH1 3LE, UK} 
\altaffiltext{5}{Kavli Institute for the Physics and Mathematics of the Universe (WPI), Todai Institutes for Advanced Study, The University of Tokyo,  Kashiwa, 277-8583, Japan}
\altaffiltext{$\dagger\dagger$}{Based on observations collected at the European Southern Observatory, La Silla Chile. Program ID 177.A-0680}

\shorttitle{ZENS IV. Similar Morphological Changes Associated with Mass- and Environment-Quenching}
\shortauthors{C.M. Carollo, A. Cibinel, S.J. Lilly et al.}

\begin{abstract}
We use the public low redshift {\it Zurich ENvironmetal Study (ZENS)} catalog to study the dependence of the quenched satellite fraction at $10^{10.0} M_\odot \rightarrow 10^{11.5} M_\odot$, and of the morphological mix of these quenched satellites, on three different environmental parameters: group halo mass, halo-centric distance and large-scale structure over-density. At a given galaxy stellar mass, the fraction of quenched satellites is more or less independent of halo mass and the surrounding large-scale structure overdensity, but increases towards the centres of the halos, as found in previous studies.  The morphological mix of these quenched satellites is, however, strikingly constant with radial position in the halo, indicating that the well-known morphology-density relation results from the increasing fraction of quenched galaxies towards  the centres of halos.  If the radial variation in the quenched fraction reflects the action of two quenching processes, one related to mass and the other to environment, then the constancy with radius of the morphological outcome suggests that both  have the same effect on the morphologies of the galaxies. The quenched satellites have larger $B/T$ and smaller half-light radii than the star-forming satellites. The bulges in quenched satellites have very similar 
luminosities and surface brightness profiles, and any mass growth of the bulges associated with quenching cannot greatly change these quantities. The differences in the light-defined $B/T$ and in the galaxy half-light radii are mostly due to differences in the  disks, which have lower luminosities in the quenched galaxies.  The difference in galaxy half-light radii between quenched and star-forming satellites is however larger than can be explained by uniformly fading the disks following quenching, and the quenched disks have smaller scale lengths than in star-forming satellites. This can be explained either by a differential fading of the disks with galaxy radius or if disks were generally smaller in the past, both of which would be expected in an inside-out disk growth scenario.  The overall conclusion is that, at least at low redshifts, the structure of massive quenched satellites at these masses is produced by processes that operate before the quenching takes place.  A comparison of our results with semi-analytic models argues for a reduction in the efficiency of  group halos in quenching their disk satellites and for mechanisms to increase the $B/T$ of low mass quenched satellites.
\end{abstract}


\section{Introduction}\label{intro}

The star formation rate (SFR) and the morphology of a galaxy are key diagnostics of its evolutionary stage. These properties are broadly correlated, in the sense that  galaxies with dominant disks tend to have higher specific SFR (sSFR=SFR/$M_{galaxy}$, with $M_{galaxy}$ the galaxy stellar mass) than galaxies with a pronounced spheroidal component.  There is solid observational evidence that the cessation of star formation in some galaxies, which results in the emergence of  `quenched'  passive galaxies, depends on both galaxy mass and environment  \citep[e.g.][]{Dressler:1980kx,Balogh:2004fk,Baldry:2006uq,Kimm:2009ij}. 
In the SDSS \citep{York:2000fk} and zCOSMOS \citep{Lilly:2007uq,Lilly:2009kx} samples, the effects of galaxy mass and environment in the quenching of galaxies appears to be separable, both in the local universe, and at least up to redshift $z=1$  (\citealt{Peng:2010fk},  Kov\^ac et al.\ 2013).  
In other words it is possible to write the fraction of galaxies that survive as blue star-forming galaxies as the product of two functions, one a function of stellar mass only, and the other a function of  `environment'  only.  Each of these functions is therefore the extra effect of mass, or environment, over the effects of the other.   In these works, the environmental measure used is an Nth-nearest neighbour estimate of the local density of galaxies.  This separability of the quenching effects of galaxy mass and environment argues has been taken to argue in favor of two independent physical processes, which \citep{Peng:2010fk} called mass-quenching (independent of environment) and environment-quenching (independent of stellar mass),  a terminology which we also adopt in this paper.   These two quenching processes seem to be quite sharp, in the sense that the sSFRs of the surviving star-forming galaxies that are not quenched are almost constant with both galaxy stellar mass and  environment (Peng et al.\  2010, 2012).

The physical nature of mass- and environment-quenching is however unclear. Several theoretical suggestions have been advanced for what we here are calling mass-quenching.   Some invoke some form of halo physics (e.g., \citealt{2007MNRAS.380..339B}, Woo et al.\ 2013, Hearin \& Watson 2013), and others include AGN feedback (see e.g., Haas et al.\ 2012 and references therein).  Other possibilities include so-called morphological quenching (\citealt{2008MNRAS.383..119D}, Martig et
al.\ 2009,2013,  Genzel et al.\ 2013).   Likewise, for environment-quenching, many physical processes have been proposed, including  ram pressure stripping, accretion shocks, and
removal of hot gas (Birnboim \& Dekel
2003, Kawata \& Mulchaey 2008, \citealt{2013arXiv1311.5916C}), with  most explanations generally invoking  some form of gas removal from galaxies as they fall into a group potential (e.g. Feldmann et al.\ 2011, De Lucia et al.\ 2012 and references therein).  

The difficulty  in identifying the mechanisms that are acting in both quenching mechanisms is at least in part due to both real and spurious correlations between different definitions of mass and environment which blur our understanding of `which mass' and `which environment' are  the most relevant for galaxy evolution. Galaxy stellar mass is directly related to dark matter halo mass, at least for central galaxies. A wide range of environmental measures on different scales may be correlated with each other, from  the global potential of the group dark matter halos, quantified by the halo mass, to the position and thus local density within the halos, to the density provided by  the large scale structure  cosmic web (e.g., \citealt{van-den-Bosch:2008fk,Kimm:2009ij,Hoyle:2011fk,peng_et_al_2012,2013MNRAS.428.3306W}; see also \citealt{2013ApJ...776...71C}, hereafter Paper I). 

One fact that has become evident, however, is that the rank of a galaxy within the group dark matter halo is very important, i.e., whether the galaxy is the first-ranked central galaxy, lying at the minimum of the potential, or a satellite galaxy orbiting within the group potential  \citep[e.g.][]{Weinmann:2006vn,Weinmann:2009fk}.  As earlier also explored by \citet{van-den-Bosch:2008fk},  \cite{peng_et_al_2012} show that the environmental effects in their 2010 analysis of the SDSS are due to changes in the satellite population alone, i.e. that their 'environment-quenching' is actually a 'satellite-quenching' process.  Within this framework, mass-quenching would be envisioned to act on all galaxies, and be related either directly or indirectly to the stellar mass of the galaxy, whereas environment- (i.e. satellite-) quenching would act only on the satellites.  Since the characteristic $M*$ of centrals and satellites is observed to be very similar, satellites must have experienced the same risk of mass-quenching as central galaxies.  It is an open question however whether this risk occurred when the satellites were actually centrals of their own halos, i.e. before they became satellites of a larger halo (see the discussion in \citealt{peng_et_al_2012}).

The same authors further showed that the fraction of satellite galaxies with red colors correlates much better with the local over-density, as estimated with a 5th-nearest-neighbour parameter, loosely interpreted to be a measure of location within the groups, than with the optical richness of the group, which is a proxy for the dark matter halo mass (see also Paper I).   Woo et al.\ (2013) also found that  the fraction of quenched galaxies at fixed galaxy stellar mass shows a  dependence on halo-centric distance only for the satellite population. These authors furthermore reported that the quenching of satellite galaxies depends also on galaxy stellar mass, but only at large (and not at small) halo-centric distances, which they interpret to arise from a quenching dependence on sub-halo mass for satellites that are recently accreted by a more massive halo.  In our picture this could be re-phrased as the increasing importance of mass-dependent mass-quenching  (affecting all galaxies) relative to mass-independent satellite-quenching  at low densities.

Important constraints for establishing the physical mechanisms behind mass-quenching and satellite-quenching can  be obtained by studying the morphologies that are associated with different populations of quenched galaxies.  It has become clear in the last few years that, in addition to the well-known connection between a quenched spectrum and an elliptical morphology, galaxies with a predominant disk component can also be quenched systems \citep[e.g.][]{Bundy:2010vn,Kovac:2010fk,Oesch:2010zr}. This prompts the question as to whether there is a dependence with environment in the morphological mix of the quenched galaxy population, and especially the quenched satellite population. The quantification of any such dependence may help elucidate in which environments quenching is or is not associated with a morphological transformation, and, more generally, which morphologies and/or morphological transformations can be associated with either mass- or environment-quenching (or both).

In this fourth paper in the  \emph{Zurich Environmental Study} (ZENS) series we use the ZENS catalog of  galactic and environmental properties published in Paper I to directly address this question by analysing the morphologies of the quenched  {\it satellite} populations that reside in a set of nearby  galaxy groups that span a range of halo masses and surrounding large-scale structure (LSS) densities.   

In ZENS  Paper I, we derived  three  measures of environment, i.e.,  the LSS over-density $\delta_{LSS}$, the mass of the host group halo $M_{GROUP}$, which we will here rename as $M_{Halo}$,  and the projected halo-centric distance $R$ relative to the characteristic radius of the group halo $R_{200}=\left(\frac{GM_{Halo}}{[10H(z)]^2}\right)^{1/3}$, which we denote $R_{vir}$ (with $H(z)=H_0\sqrt{\Omega_M(1+z)^3+\Omega_{\Lambda}}$ the Hubble constant at the given redshift).

A strength of ZENS is to  eliminate  sample-dependent biases by using exactly the same sample of galaxies to perform such environment-vs-environment comparisons.  These galaxies are 1455 members of 141 groups which were extracted from the Percolation-Inferred Galaxy Group, 2PIGG, catalog, \citealt{Eke:2004qf}, of the 2dFGRS, \citealt{Colless:2001ys}).

The present  paper explores specifically, at fixed galaxy stellar mass; $(i)$ how the fraction of quenched satellites depends on group halo mass, on halo-centric distance and on large-scale structure density,  and $(ii)$ how the morphological composition of the quenched satellite populations depend on these different environmental measures.   Indirect estimates for the quenched fraction of the central galaxies of similar stellar mass are also discussed, in order to obtain a benchmark estimate  for the relative contribution of  mass-quenching and satellite-quenching at a given galaxy mass scale.

The paper is organised as follows. In  Section \ref{survey}  we briefly describe ZENS, and in particular  the  environmental parameters derived in Paper I  as well as the morphological and spectral  parameters for the ZENS galaxies that are  derived respectively in \citet{2013ApJ...776...72C}, hereafter Paper II,  and \citet{2013ApJ...777..116C}, hereafter Paper III\addtocounter{footnote}{-7}\footnote{The full ZENS catalog of galaxy and environmental measurements is  published  with Paper I.}.  In Appendix \ref{allrelaxed} we show the impact of including  unrelaxed groups in the present  analysis, and define the ZENS sample that we use in the present study, which includes both relaxed and unrelaxed groups (see Section \ref{cleanstuff} for the definitions). 
In Section \ref{halosize}, we present the dependence on $M_{Halo}$ and $R/R_{vir}$ of the total fraction of quenched satellites  $f_Q$,  and  of the fraction $f_{Q_{ETG}}$ of quenched satellites that have an early-type galaxy (ETG) morphology, relative to the total quenched satellite population.  In Section \ref{lss},  we  separately report the analysis of the satellite quenched fraction and its morphological composition as a function of LSS density. In  Section  \ref{discuss} we investigate the global properties of quenched and star-forming satellites, and of their bulges and disks, and discuss our results first in terms of the apparent morphological changes associated with mass- and environment-quenching, and secondly in comparison  with  traditional semi-analytic models (SAMs) of galaxy formation.  Section \ref{con} summarises our findings and highlights some open questions that emerge from our analysis. 

All our ZENS work is based on a $\Omega_m=0.3$, $\Omega_{\Lambda}=0.7$ and $h=0.7$ cosmological model. 

\section{Essentials on ZENS}\label{survey}

Details on the ZENS sample, multi-band data,  environmental measurements,   and  measurements of galaxy  morphology, structure, stellar mass,  and star formation rate have been described in  Papers I, II and III. Briefly, the ZENS group sample  spans the halo mass range from $\sim 10^{12.2} \mbox{M}_{\odot}$ to $\sim 10^{14.9} \mbox{M}_\odot$.  Halo masses were determined from the group luminosities.  SAMs were used to estimate the intrinsic uncertainty  in the conversion between halo luminous and dark masses, and tests were performed to assess other systematic uncertainties, in particular the impact of interlopers and missed galaxies. Due to these uncertainties, scattering of groups between the two halo mass bins, i.e. cross-talk between halo mass bins, may weaken any intrinsic dependence of galaxy properties on halo masses, a factor which we will take into account when analysing our results. The ZENS group sample  was selected to be in the narrow redshift range $0.05 < z < 0.0585$ and to include groups with at least 5 spectroscopically-confirmed member galaxies down to the limiting magnitude ($b_J=19.45$ mag) of the 2dFGRS.   We highlight  below those aspects of ZENS that are most relevant to the current analysis.

\subsection{Definition of the different environments and minimisation of environmental cross-talk}

In computing different environmental parameters, which in principle should probe the  surroundings of galaxies on different scales, several factors can introduce  cross-talk amongst such  parameters. This complicates attempts to  identify  the physical mechanisms behind environmentally-driven galaxy evolution. We discuss below  some of  the most important of these factors, which are relevant for our analysis, and summarise our approaches to minimise their impact on the results.

\subsubsection{Identification of centrals and satellites, and definition of  relaxed and unrelaxed groups}\label{cleanstuff}
 
 Accurate knowledge of group centres is crucial for reliably measuring  variations of galaxy properties with halo-centric distances and  differentiating central galaxies from satellite galaxies. Identifying the central galaxies and thus the group centres can however be difficult, especially for groups with few spectroscopically confirmed galaxy members.  
In optically-selected galaxy group catalogs, the customary approach to identifying the central galaxy is to find the most massive, or sometimes even just the brightest, galaxy  amongst those that are linked as a single group by a Friend-of-Friend (FoF) algorithm.  This leads in some cases to  groups in which the identified central galaxy is substantially displaced from the group centres, either in projected distance or in velocity space.  

To address  this issue, in ZENS we classified groups as `relaxed' and  `unrelaxed', depending on whether we could identify a single central galaxy that simultaneously satisfies the mass, halo-centric distance and velocity requirements for it to be assigned the rank of central galaxy.   Specifically, we initially identified the apparently most massive galaxy  (i.e., the member galaxy with the highest best-fit stellar mass, as established from synthetic galaxy template fitting to the observed  NUV-to-NIR photometric Spectral Energy Distributions [SEDs]).  Groups were classified as relaxed  if this most massive galaxy was located  (i) within a radial distance of 0.5$R_{vir}$ from  the mass-weighted geometric center of the group, and (ii) within the central 68 percentile of the velocity distribution of the group.  Groups in which the highest mass galaxy did not satisfy these space and velocity criteria were further studied to assess whether, within the errors of our galaxy stellar mass estimates, there was a different  galaxy which satisfied all the above-mentioned criteria within the errors. Groups for  which this procedure  found an alternative bona-fide central galaxy  were also classified as relaxed (but are excluded from the `clean' sample of relaxed groups to which we refer below). Groups in which no self-consistent solution could be found for a bona-fide central, i.e., a galaxy that, within the errors, satisfied all the mass, location, and velocity  criteria, were classified as unrelaxed.  The ZENS sample contains 59 unrelaxed    groups and 82 relaxed  groups (73 of which compose the `clean' sample of relaxed groups in which the central galaxy is the  nominally most massive galaxy).

\subsubsection{An LSS density estimate unaffected by group richness }\label{lssnew}

As pointed out by several authors, computations of the LSS density based on Nth nearest-galaxy approaches correspond to different environments in halos of richness above or below  the chosen `N', describing respectively inter- and intra-halo densities (Peng et al.\ 2010, \citealt{2013MNRAS.428.3306W}, Paper I). In ZENS we adopted a parameterization of the LSS density which does not switch meaning for galaxies in groups of richness above and below a given value and self-consistently measures the inter-group density of the filamentary cosmic structure. Specifically, we traced the  LSS  density using  a 5th-nearest-neighbor approach, adopting however the mass-weighted {\it groups} rather than their member {\it galaxies} as the tracers of the density field.  All groups in the 2PIGG catalog within the redshift interval $dz = \pm 0.01$ of the group of interest were used in this computation, including `one-galaxy groups',  i.e. isolated galaxies that are more luminous than a $z=0.07$,   $b_j=19.1$ mag   galaxy. 
Our LSS density parameter thus reflects the location of a group in the cosmic web and not at all the local density that a given satellite will experience within its group halo since, by definition, all galaxies members of a given group are assigned an identical LSS density parameter.

\subsubsection{Disentangling  halo mass and LSS density effects} 

The physical association of high mass halos with high LSS densities introduces a degeneracy that makes it difficult to disentangle the separate effects of physics within virialized massive halos (as parametrized by the halo mass) and dense regions of the cosmic web (as parametrized by the inter-halo LSS density). To attempt to establish whether possible environmental trends are driven by one or the other of these physically distinct environments, we  apply a high halo-mass cutoff in the group sample when  investigating possible effects of the LSS density on galaxy properties. Specifically, we  use only groups with  $M_{Halo}\leq10^{13.7}\ \som$ in Section \ref{lss} to study whether the surrounding LSS density affects either the quenched satellite fraction or its morphological composition. These low-to-intermediate mass  halos can be found over a wide range of LSS densities (see Paper I).  

\subsection{Identification of  the quenched galaxies}

To separate galaxies into quenched and (moderately or strongly) star-forming systems, we adopted the combination of several indicators.  Quenched galaxies are required to satisfy all of the following spectral and color criteria:  (i) No detected emission in H$\alpha$ and H$\beta$, and  (ii) $(NUV-I) > 4.8$, $(NUV-B) > 3.5$ and $(B-I) > 1.2$. 

The final assignment of a  galaxy to the quenched  population was based on the sSFR derived from an SED fit to its photometric data (see Section \ref{msfr}), validated however through the spectral and color requirements listed above. This multiple-constraint approach to the identification of quenched galaxies leads to a  higher purity of the sample compared to the use of a single (often optical) color criterion; samples based on the latter may suffer from  a non-negligible contamination from, for example, dust-reddened star-forming galaxies, especially at masses $\lta10^{11} M_\odot$ (see Paper III  and  \citealt{2013MNRAS.428.3306W}).  

In discussing our results in Section \ref{discuss}, we will compare the bulge and disk properties of the quenched satellites with those of  star-forming satellites  of similar stellar mass in the ZENS sample.  In our work we define galaxies to be strongly star-forming if they show: $(i)$ strong  [OII], [NII],  H$\alpha$ and H$\beta$ line emission,  as well as $(ii)$ $(NUV-I) < 3.2$, $(NUV-B) < 2.3$ and $(B-I) < 1.0$ colors.   Galaxies with color and spectral properties in-between strongly star-forming and quenched galaxies are classified as moderately star-forming objects. Further details are given in Paper III.

\subsection{Estimates of  galaxy stellar masses and SFRs}\label{msfr}

Galaxy stellar masses and SFRs (and thus the sSFR values used in the identification of quenched galaxies) were derived in Paper III by fitting model SEDs to the photometric data of the quenched galaxies using the {\it Zurich Extragalactic Bayesian Redshift Analyzer+} (ZEBRA+; \citealt{Oesch:2010zr}). ZEBRA+  is an upgraded version of our publicly-released ZEBRA code \citep{Feldmann:2006ly}. Stellar population models were  adopted from the \cite{Bruzual:2003kx} library with a Chabrier (2003) initial mass function. We used two types of star formation histories, i.e. exponentially decaying models with a broad range of characteristic $\tau$ timescales (from very short to very long e-folding $\tau$ values), and constant star formation models.  Details on the grid of SED models are given in Table 2 of Paper III.

For galaxies with clearly quenched (or star-forming) spectra and colors, initial discrepancies between SED-based SFRs and the spectral+color assessments were individually resolved by performing a second SED fit, after restricting the sample of  templates  to quenched (star-forming) SED models only. 

We use as our definition of  stellar mass the integral of the past star-formation rate. Our stellar masses are thus about 0.2 dex higher than the `actual' stellar mass that is sometimes used, i.e., the stellar mass that remains after mass-return to the interstellar medium from the evolving stellar population.  As a result, the sSFRs in this paper are the inverse mass-doubling time scale.

With our definition, the stellar mass completeness of the ZENS galaxy sample for quenched systems is  $M_{galaxy}\sim 10^{10}{M}_\odot$. 

\subsection{Galaxy morphologies and $B/T$ decompositions}\label{morphs}

A strength of ZENS is its correction for measurement biases in structural parameters.  These were calibrated to eliminate spurious trends with the size of the seeing-driven Point Spread Function (PSF), and with galaxy magnitude, size, concentration and ellipticity.  A quantitative morphological classification was then implemented on the basis of these corrected structural parameters. The classification was mainly based on the light-defined bulge-to-total ($B/T$) ratio, which was measured via two-dimensional bulge+disk fits to the galaxy surface brightness distributions, reinforced by imposing quantitative boundaries for the different morphological types in standard non-parametric quantities such as concentration, Gini, $M_{20}$ and smoothness parameters (see Table 2 and Figure 18 of Paper II for details).   Bulges were modelled with a S\'ersic   profile and disks with a single exponential profile (see Paper II for details).

For the present analysis, we divide galaxies into three morphological bins, based on the basic ZENS morphological classes described in Paper II. These three bins are here referred to as the early-type, disk-dominated and bulgeless disk galaxies. With reference to the morphological classes defined in Paper II,  the early-type bin  includes both elliptical galaxies, i.e., galaxies described by one single component with S\'ersic index $n>3$, and bulge-dominated disks with B/T$\geq 0.5$. The disk-dominated bin is composed of  intermediate-type disks with $0.2\leq$B/T$<0.5$, i.e., of disk galaxies with a substantial but not dominant bulge component, and the bulgeless bin consists of late-type disks with B/T$<0.2$.  

We note here that, at the galaxy mass scales of  this study, bulgeless disks contribute negligibly to the morphological mix of the {\it quenched}  galaxy population. Therefore, in the following we will simplify the presentation of our results by reporting only the fractional weight $f_{Q_{ETG}}$ of the quenched early-type satellite population, relative to the whole quenched satellite population of similar mass. Given the paucity of bulgeless disks within  the $M_{galaxy} \geq 10^{10} M_\odot$ quenched satellite population, a value of $(1-f_{Q_{ETG}})$ gives, in practice, the contribution of  the aforementioned disk-dominated galaxies to the total quenched satellite population.

\subsection{The final sample for the present analysis}

The completeness mass limit of $10^{10} M_\odot$ for quenched galaxies motivates our adoption of this value as the lower galaxy mass limit in this study. We thus  conduct our analyses  in two well-defined bins of galaxy stellar mass: a low galaxy stellar mass bin  defined by  $10^{10}{M}_\odot \leq M_{galaxy}\leq 10^{10.7}{M}_\odot$, and  a high mass bin, $10^{10.7}{M}_\odot \leq M_{galaxy}\leq 10^{11.5}{M}_\odot$.     The median values of our galaxy samples in these two mass bins are  respectively $10^{10.32}$ M$_{\odot}$ and $10^{10.85}$ M$_{\odot}$.    The explored range of galaxy stellar mass straddles the characteristic (and seemingly invariant) value of $M*$ of the \cite{Schechter:1976ve} function that describes both the quenched and star-forming  galaxy populations. As discussed in \citet{Peng:2010fk}, $M*$  emerges as the characteristic mass in the quenching of galaxies. With our convention for defining stellar masses,  $M* \sim 10^{10.85} M_\odot$, which matches the median value of our galaxy sample in the  high galaxy stellar mass bin.  

As shown in Appendix \ref{allrelaxed}, the fraction of quenched satellites and morphological mix of these quenched satellites remain practically unchanged, in any environment,  with the inclusion or exclusion of the unrelaxed groups in the computations. To maximise the size of the different galaxy samples, we will therefore use the entire  ZENS  group sample, independent of the group dynamical state, in studying  the  dependence   on the different environments of the quenched satellite fraction and its morphological composition (Sections \ref{halosize} and \ref{lss}).

\begin{subfigures}
\begin{figure*}[t!]
\begin{center}
\includegraphics[width=0.9\textwidth]{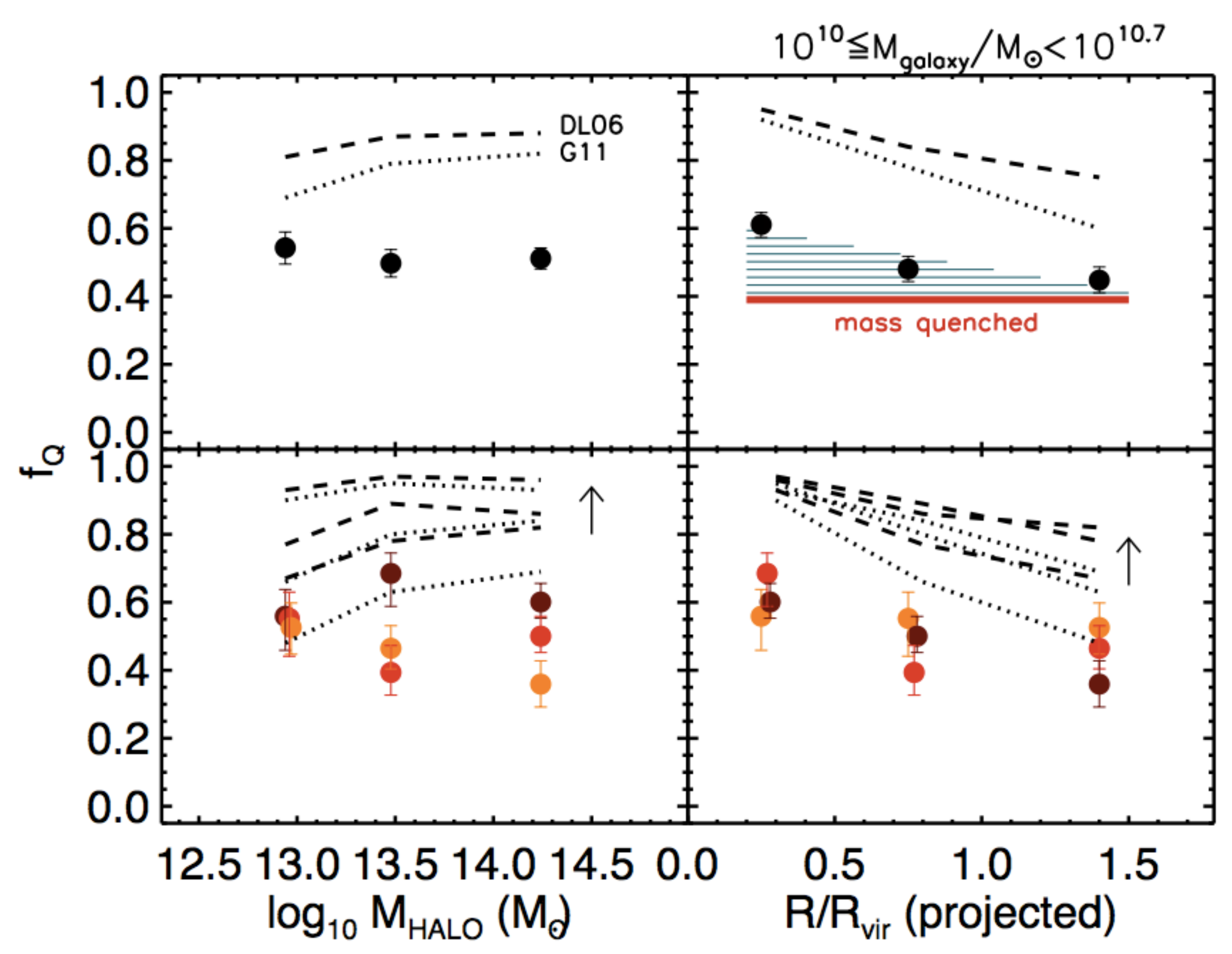}
\caption[]{ Fraction $f_Q$ of quenched satellite galaxies of all morphologies with stellar mass in the range $10^{10-10.7}M_{\odot}$,
as a function of halo mass (left) and  halo-centric distance (right). Both relaxed and unrelaxed groups are included in the sample (see discussion in Appendix  \ref{allrelaxed}).
The error bars indicate Bayesian $\pm 1 \sigma$ confidence intervals for a binomial distribution.
In the upper panels, satellites at all halo centric distances (left) or at all halo masses (right) are included in the sample. The red horizontal line in the top-right panel indicates the level of mass-quenching, and the hatched area above the red horizontal line highlights the quenching-excess attributed to environmental-quenching. 
 The bottom panels show the quenched fractions of satellite galaxies, this time split over the bins in the  second environmental parameter which are defined by the dotted lines of Figure \ref{fA1} (i.e., bins of $R/R_{vir}$ for the plot of $f_Q$ versus $M_{Halo}$, and  bins of $M_{Halo}$ for the plot of  $f_Q$ versus $R/R_{vir}$). Colors range from light (yellow) to dark (brown) moving from sparse  (low $M_{Halo}$, high $R/R_{vir}$) to dense   (high $M_{Halo}$, low $R/R_{vir}$) environments.
The black lines show the predictions of the semi-analytic models of \citealt{De-Lucia:2006zr} (DL06, dashed line) and   \citealt{Guo:2011ve} (G11, dotted line). In the bottom panels, the models are split in bins of the second environmental parameter, as done for the ZENS  data points. The black arrows indicate the direction of  increase in density for the models, from sparse to dense environments.\label{f1a}
}
\end{center}
\end{figure*}

\begin{figure*} 
\begin{center}
\includegraphics[width=0.9\textwidth]{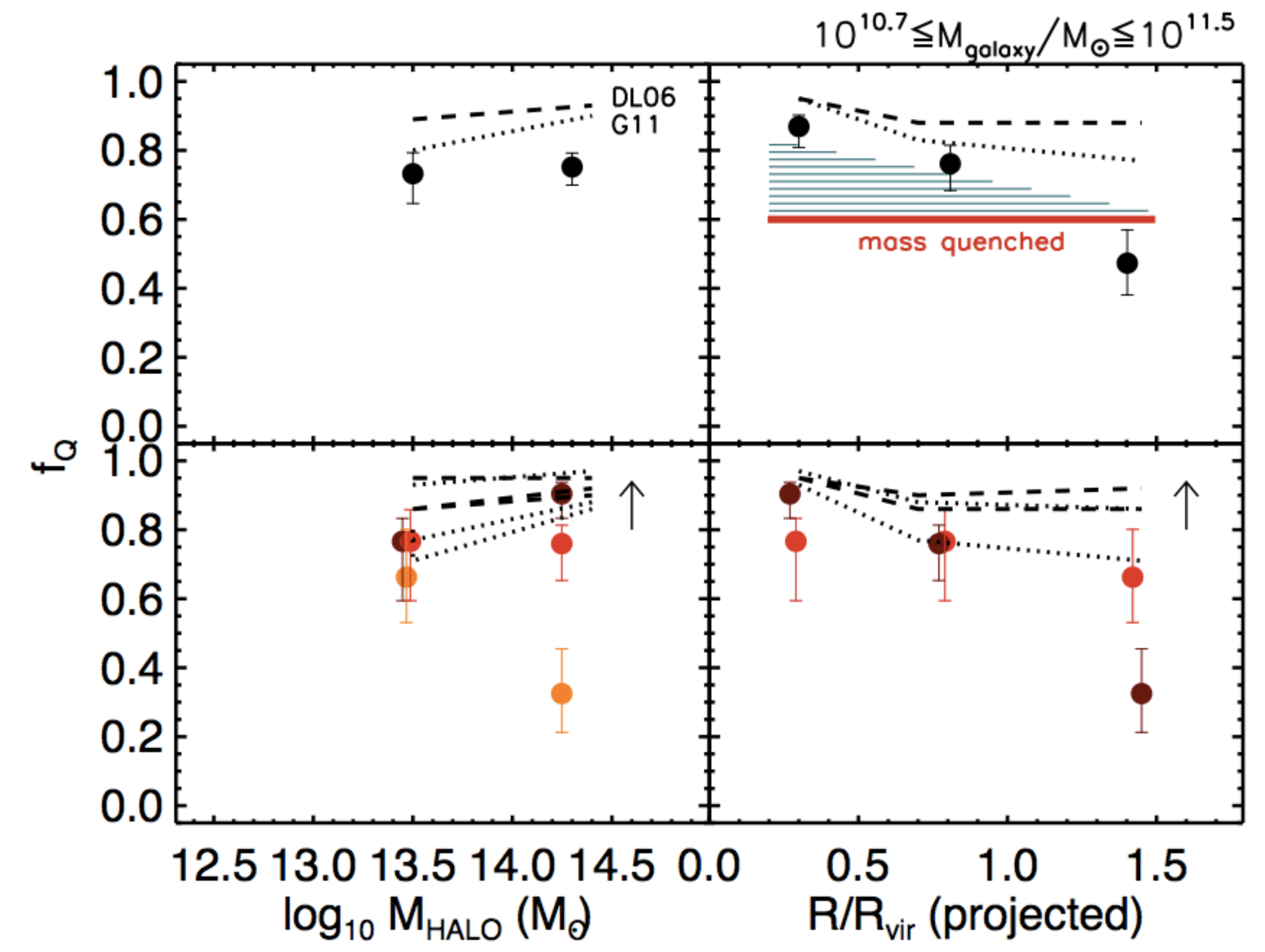}
\caption[]{
As in Figure \ref{f1a}, but for the galaxy stellar mass bin $10^{10.7}-10^{11.5}M_{\odot}$.  \label{f1b}}
\end{center}
\end{figure*}
\end{subfigures}

\section{Quenching within halos}\label{halosize}

In Table \ref{t1}   we list the fraction $f_Q$ of quenched satellites, and the fraction $f_{Q,ETG}$ of quenched satellites that have an early-type morphology, in different environmental bins of  the $M_{Halo}$-$R$ parameter space. Specifically we computed these fractions  in nine and six environmental bins for the low and high galaxy stellar mass bins, respectively.   The nine environmental bins at  low galaxy stellar mass are defined by $M_{Halo} = [12.2,13.2[ , [13.2,13.7[$ and  $[13.7,14.8]$  and  $R\leqslant 0.5R_{vir}$, $ 0.5R_{vir}<R \leqslant  R_{vir}$ and $R > R_{vir}$; the six environmental bins at high stellar mass have a similar split in $R$ but  a two-bin split in halo mass, i.e., $M_{Halo} = [12.8,13.7[$  and  $[13.7,14.8]$ (since, as a consequence of the sample selection criteria, there are no such massive galaxies in ZENS at lower halo masses). These environmental bins are    illustrated, for the low and high stellar mass bins, respectively in the left and right panels of Figures \ref{fA1} and \ref{fA2}. To show the insensitivity of the quenched fractions and their morphological mix  to the precise group sample selection,  the Table reports the fractions that we obtain  by using different group samples, i.e.,  all relaxed groups, the `clean' sample of  relaxed groups (see Section \ref{cleanstuff}), and the full ZENS group sample which includes also the unrelaxed groups. As discussed in Section \ref{survey} and Appendix \ref{allrelaxed}, we base the following analysis on the entire, relaxed plus unrelaxed  group  sample.

\subsection{The dependence of the quenched satellite fraction on halo mass and halo-centric distance}\label{quenchedfraction}

Figures \ref{f1a} to \ref{f2b} show the results of Table \ref{t1} in graphic form. Specifically, the upper panels of Figure \ref{f1a} show the quenched fraction $f_Q$ of our lower mass $10^{10-10.7} M_\odot$ satellite galaxies as a function of group halo mass $M_{Halo}$ (upper right) and normalized halo-centric radius $R/R_{vir}$ (upper left).   The lower panels show the same results split however into the three bins of halo-centric radius (lower left plot) and for the three bins of halo mass (lower right plot) that are listed above and in Table \ref{t1} (and illustrated by the dotted lines in Figure \ref{fA1}).   Figure \ref{f1b} shows the same for the higher mass satellite galaxies ($10^{10.7-11.5} M_\odot$), but now, as discussed above and in Appendix \ref{allrelaxed}, we split the sample in only two bins of halo mass in the lower left panel.  In both Figures, we plot the median quenched fractions in the corresponding environmental bins, i.e., the values reported in Table \ref{t1},  at the median values of the environmental parameters within the bins in question.

At the galaxy mass scales that we are probing, we find no significant dependence on halo mass of the quenched satellite fraction $f_Q$, integrated over all halo-centric distances: about 50$\%$ and 70\% of  $10^{10-10.7} M_\odot$ and $10^{10.7-11.5} M_\odot$ satellite galaxies, respectively, are quenched systems at all halo masses.  Bootstrapping  simulations give a $1\sigma$ upper-limit slope $\frac{df_{Q}}{dlog M} \leq 0.1$ for  the $f_Q$ versus $M_{Halo}$ relation across the range covered by our data. 

In Paper I, we estimated a typical uncertainty of about 0.3 dex for the  halo masses, which in general would tend to weaken any actual dependence of parameters with halo mass, even though the width of the halo mass bins in our analysis is larger than this typical uncertainty. Mock simulations, discussed in Paper I, indicate that the observed flat relation between $f_Q$ and $M_{Halo}$ does indeed imply a  negligible intrinsic slope for this relationship.   In their SDSS analysis, Woo et al.\ (2013) find a rather weak positive correlation between their satellite red fraction and halo mass, while Peng et al.\ (2012) find  that  the fraction of quenched satellites is invariant with halo mass (at fixed local overdensity), as we do in ZENS;   these latter authors interpret their result as indicating that satellite-quenching does not depend on halo mass. 

In contrast to the apparent independence with halo mass, and in agreement with previous independent analyses, we find a significant  environmental dependence of $f_Q$ on halo-centric distance $R$: the quenched fraction, averaged over all halo masses,  decreases with increasing  $R$ (see upper-right plots of   Figures  \ref{f1a} and \ref{f1b}). This is also seen in the  SDSS analysis of Woo et al.\, and, indirectly,  Peng et al.\ (2012) if their local projected density $\rho$ is interpreted as a proxy for the halo-centric radius; other studies have also found a similar trend  (e.g., Rasmussen et al.\ 2012, \citealt{2013ApJ...770..113G}).  In our data, this radial trend is seen mostly, and most strongly, at the highest halo masses.  When averaged over all halo mass scales, the fraction $f_Q$ of low mass quenched satellites in our sample declines from   $\gta 60 \%$ in the cores of groups, down to $\lta 50\%$ at the virial radius, whereas the high mass satellites decrease from $\gta 80\%$ down to $\lta 60\%$). The total quenched fraction keeps decreasing with increasing $R/R_{vir}$, marginally in the low galaxy mass bin and significantly in the high galaxy mass bin, where our $R>R_{vir}$ measurement  gives  $f_Q\sim50\%$. 

We also note that the general trends  that we have highlighted for the total quenched fraction $f_Q$ may break down at  large halo-centric distances $R>R_{vir}$ and towards low halo masses.  A concern remains that, at least in ZENS, these sub-samples are small and most satellites at very large halo-centric distances are members of unrelaxed groups (although the effects appears to persist independent of whether the unrelaxed groups are considered in the analysis).  These possible breakdowns of the more global trends should be explored on larger samples.

 \subsection{The dependence of the morphological mix of the quenched satellite fraction on halo mass  and halo-centric distance}\label{morphologicalmixpsf}

In Figures \ref{f2a} and \ref{f2b} we show the morphological composition of the quenched satellite population, for the two bins of satellite stellar mass, as functions of halo mass and halo-centric radius.   Specifically, we plot here the fraction of quenched satellites that have early-type morphology, $f_{Q_{ETG}}$.   As commented in Section \ref{morphs}, there are virtually no bulgeless quenched galaxies at the galaxy mass scale that we are studying, and the remaining quenched satellites are therefore intermediate-type disks with a dominant disk component but a non-negligible bulge.

We find no trends at all in the morphological mix of the quenched satellite population with either halo mass or halo-centric radius, despite the significant radial trends seen in the total fraction of quenched satellites  in Figures \ref{f1a} and \ref{f1b}.   All of the data points in Figures \ref{f2a} and \ref{f2b} are consistent with a constant early-type fraction of about $50\%$ for $10^{10-10.7} M_\odot$ satellite galaxies, and about 65\% for the more massive $10^{10.7-11.5} M_\odot$ ones.  Thus, while it is clear from Figures \ref{f1a} and \ref{f1b}  that more satellites have been quenched in the cores of groups than in their outskirts, the {\it relative} fractions  of early-type and disk-dominated morphologies in the quenched satellite population does not change with group-centric distance.

The striking constancy of the $f_{Q_{ETG}}$ with halo-centric radius in Figures \ref{f2a} and \ref{f2b} does not of course violate the well-known morphology-density relation of Dressler (1980).  Our results  indicate  that this is due to the changing fraction of quenched galaxies with radius $f_Q(R)$, rather than a change in the morphologies of the quenched galaxies themselves (we show below that there is a systematic difference between the morphologies of the quenched satellites and their star-forming counterparts).

\begin{subfigures}
\begin{figure*}  \begin{center}
\includegraphics[width=0.9\textwidth]{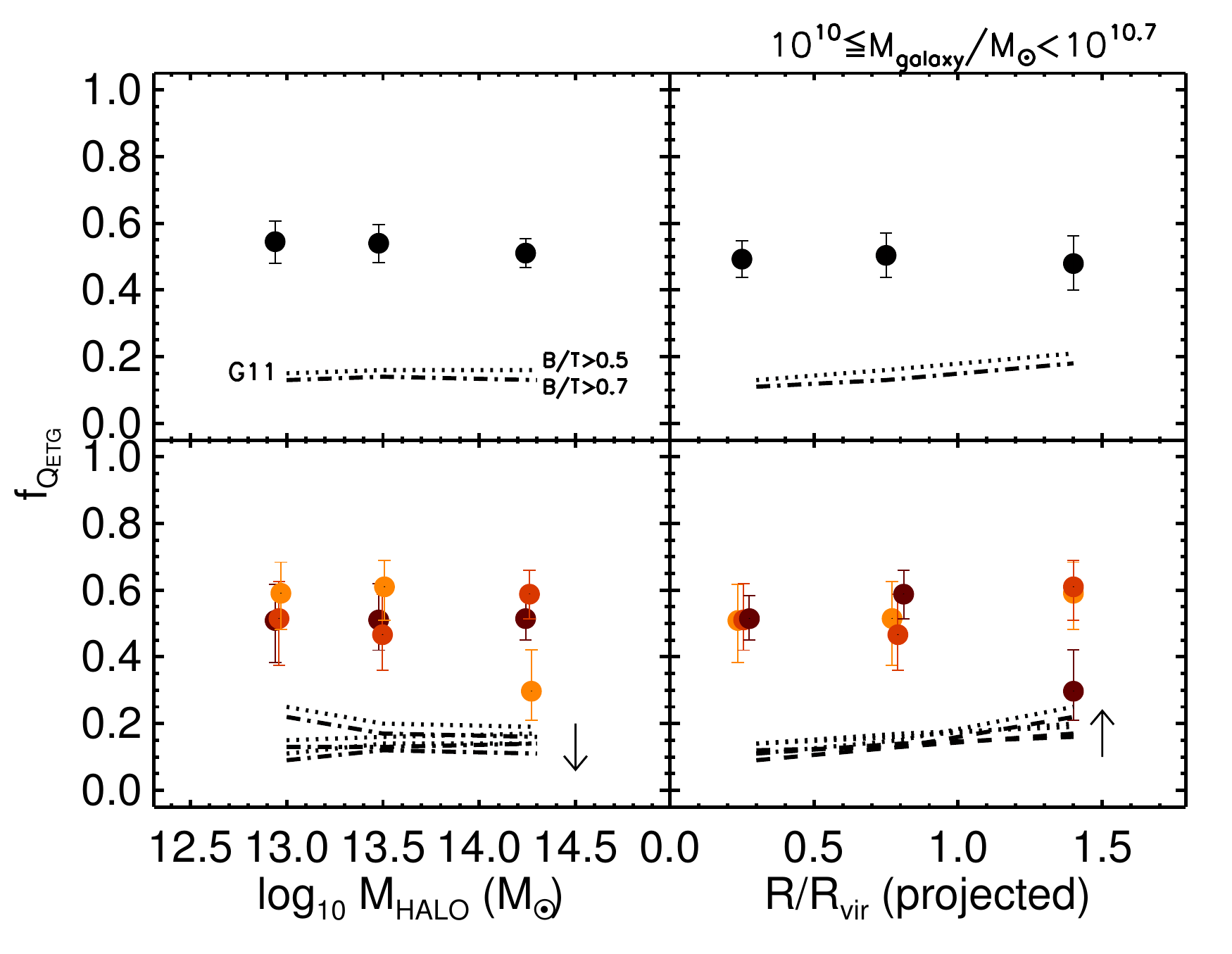}
\caption[]{Fraction of quenched satellite galaxies with an early-type morphology as a function of group halo mass $M_{Halo}$ (left panel) and halo-centric distance $R/R_{vir}$ (right panel), in the galaxy stellar mass bin $10^{10}-10^{10.7} M_\odot$. Both relaxed and unrelaxed groups are included in the sample (see discussion in Appendix \ref{allrelaxed}). In the upper panels, satellites at all halo centric distances (left) and at all halo masses (right) are included in the sample.
The bottom panels show again the quenched fractions of satellite galaxies with an early type morphology  relative to  $f_Q$, this time   split over the  bins in the second environmental parameter which are defined by the dotted lines of Figure \ref{fA2} (i.e.,  bins of $R/R_{vir}$ for the plot of $f_{Q_{ETG}}$ versus $M_{Halo}$, and  bins of $M_{Halo}$ for the plot of  $f_{Q_{ETG}}$   versus $R/R_{vir}$). Colors range from light (yellow) to dark (brown) moving from sparse  (low $M_{Halo}$, high $R/R_{vir}$) to dense   (high $M_{Halo}$, low $R/R_{vir}$) environments.
 The black lines show the predictions of the semi-analytic models of  \citealt{Guo:2011ve} (G11). Dashed-dotted lines indicate model early-type satellite galaxies defined as systems with a $B/T>0.7$ in stellar mass; the dotted lines are for model satellites with in stellar mass of $B/T>0.5$. The models in the bottom panels are split in   bins of the second environmental parameter, as done for the ZENS data. The black arrows indicate the direction of  increase in density for the models, from sparse to dense environments. \label{f2a}}
 \end{center}
\end{figure*}

 \begin{figure*}  \begin{center}
 \includegraphics[width=0.9\textwidth]{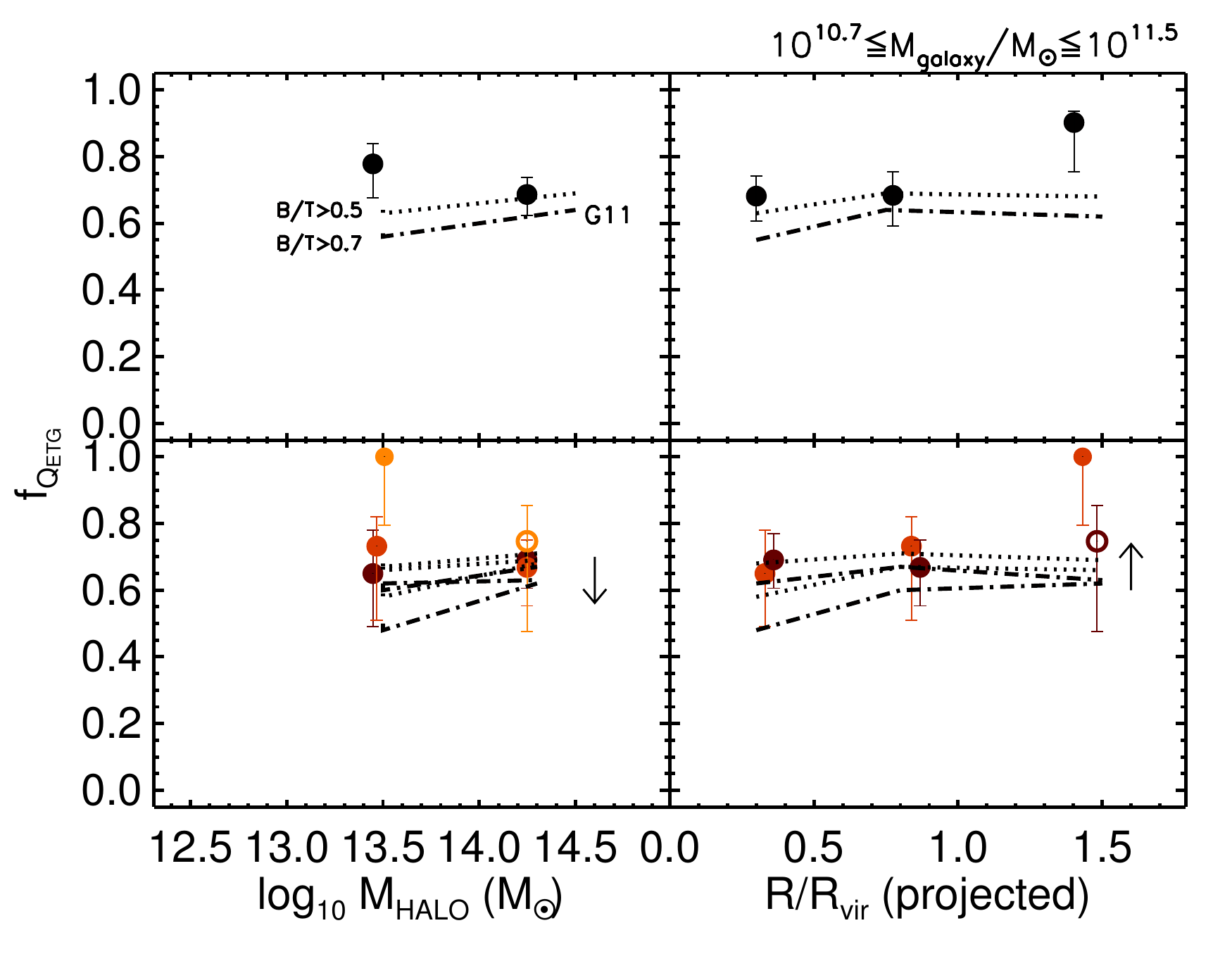}
 \caption[]{As for Figure \ref{f2a} but for the galaxy stellar mass bin $10^{10.7}-10^{11.5} M_\odot$. In the lower panels, open symbols indicate  $\leq5$ objects at the denominator of the given fraction.\label{f2b}}
 \end{center}
\end{figure*}
\end{subfigures}

\section{Influences on quenching from outside of halos?}\label{lss}

We can ask whether quenching processes and possible morphological transformations may be triggered by the average LSS density on which the group halos reside, and thus search  for trends of $f_Q$ and $f_{Q_{ETG}}$ with the average LSS density parameter $\delta_{LSS}$ discussed in Section \ref{lssnew}. We recall that this parameter represents the average density at the location of a given group, and is identical for all galaxies that are members of that group. Also, in order to minimize as much as possible the physical cross-talk between high LSS densities and high halo masses, we exclude the most massive halos  in our sample for this analysis. Specifically, as in Paper I and as noted above, we use for this analysis only groups with $M_{Halo} \leq 10^{13.7} M_\odot$. As before, we use here both relaxed plus unrelaxed  groups satisfying this mass condition.

Taking into account the range of $\delta_{LSS}$ covered by the galaxy samples, we split the LSS density domain into three bins with boundaries at 
$\log_{10}(1+\delta_{LSS}) = 0.3$ and 0.7 for the low mass sample and into two bins split at $\log_{10}(1+\delta_{LSS}) = 0.7$ for the high mass sample.
We then compute the fractions of quenched satellites within these boundaries of LSS density in the same three bins of halo-centric distance already defined in Section \ref{halosize}, i.e., $R\leqslant 0.5R_{vir}$, $ 0.5R_{vir}<R \leqslant  R_{vir}$ and $R>R_{vir}$. The  nine (six) environmental bin splitting of  the $\delta_{LSS}-R$ plane at low (high) galaxy stellar mass are  illustrated  in the left (right) panel of Figure \ref{fA3} of Appendix \ref{allrelaxed};   the corresponding quenched and quenched-ETG fractions are listed in  Table \ref{t2}.

We  show in Figures \ref{f4a}-\ref{f4b} the analogous plots to Figures \ref{f1a}-\ref{f2b}, this time however using  the underlying density of the cosmic LSS as the environmental parameter on the x-axis.  At high galaxy masses, our $M_{Halo} \leq 10^{13.7} M_\odot$ ZENS  sample    is  too  small to enable us a split  in different bins of  halo-centric distance (see the right panels of Figures \ref{fA3} and \ref{fA4}). Thus Figures \ref{f3b} and \ref{f4b} are analogous plots only to the upper panels of Figures \ref{f3a} and \ref{f4a}. Formally we observe an increase of $f_Q$ with increasing LSS density, which interestingly well matches the trend  predicted by, for example, the Guo et al.\ (2011) semi-analytical model (which we discuss further in Section \ref{mod};   dotted line in Figure \ref{f3b}).  In contrast, the decrease of $f_{Q_{ETG}}$ with increasing $\delta_{LSS}$  is opposite to that predicted by this same model (dotted and dash-dotted lines in Figure \ref{f4b}).  Given however that, at these high masses, even in the averaged (over R) relationships of Figures \ref{f3b} and \ref{f4b} the statistical uncertainties are substantial and thus the significance of the trends observed with $\delta_{LSS}$ is low, we will not include them in the discussion of our  main results in Section \ref{discuss}. 

In the lower galaxy mass bin we have, in contrast, enough statistics to properly explore the dependence of the quenched fraction and its morphological mix on the LSS density. We find no evidence for a relationship between  $f_Q$ or $f_{Q_{ETG}}$ and $\log (1+\delta_{LSS})$ , and thus  no evidence  of any impact of the LSS density at which the groups reside on the fraction  of  $10^{10-10.7} M_\odot$  quenched satellites that inhabits  these groups (at any halo-centric distance, see bottom panel  of Figure \ref{f3a}). This means that, at least at these galaxy masses, the quenching of satellites within a halo does not   depend on the location of that halo in the filamentary cosmic web.   Also the morphological mix of the quenched fraction at this galaxy mass scale is independent of LSS density surrounding the halos in which these satellites reside. Again, this  is  true regardless of  the halo-centric distance of the satellite within the halo (see bottom panel of Figure \ref{f4a}).

\begin{subfigures}
\begin{figure} \begin{center}
\includegraphics[width=0.45\textwidth]{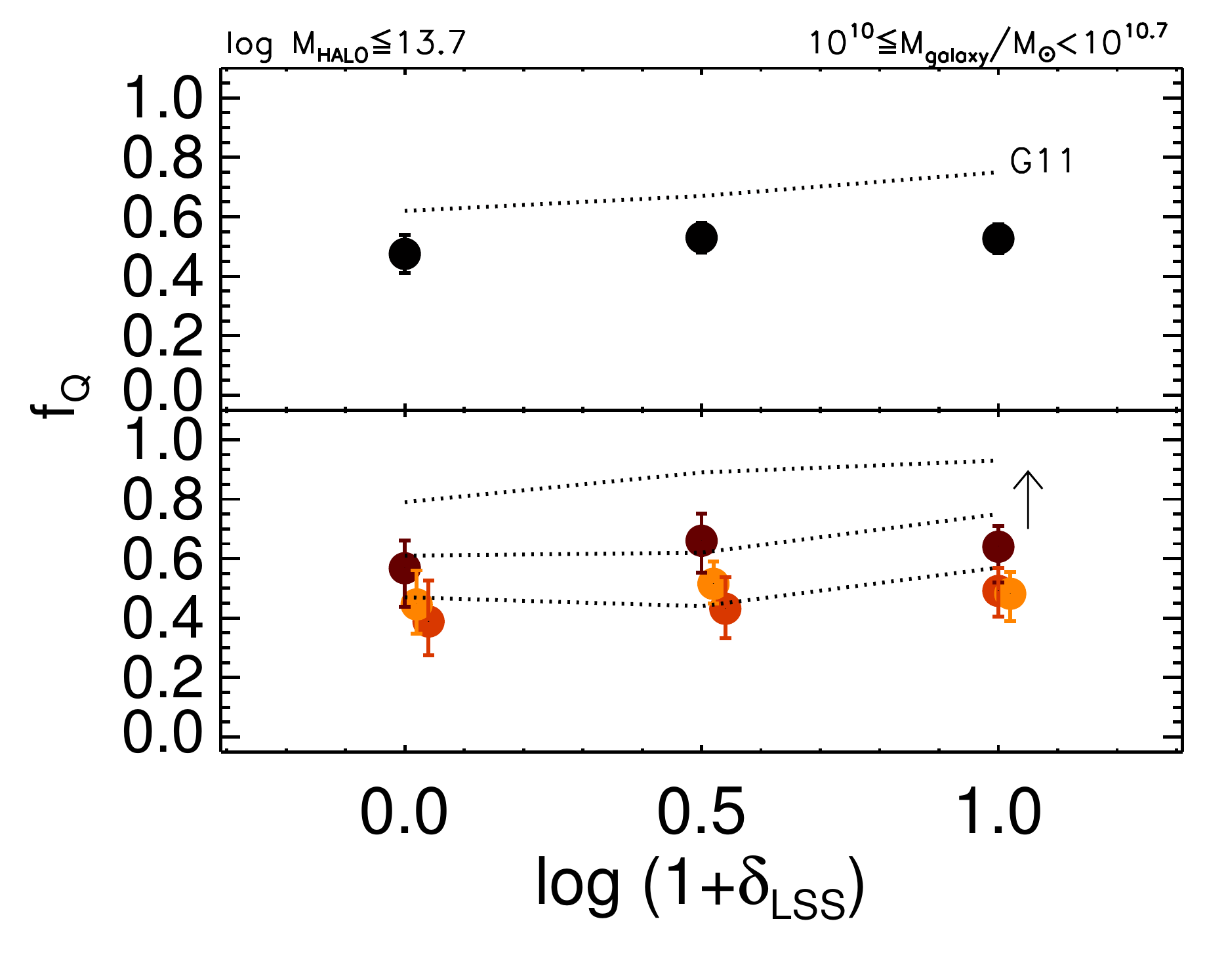}
\caption[]{The fraction $f_Q$ of  quenched satellite galaxies (of any morphology) as a function of  LSS density, for the galaxy stellar mass bin $10^{10}-10^{10.7}   M_\odot$.  Our LSS density parameter  reflects the location of a group in the cosmic web (not the local density of a given satellite within its group halo). The top panel shows the total fraction integrated over halo-centric distance; the bottom panel shows the fractions in each of the  bins of $R/R_{vir}$ highlighted  in Figure \ref{fA3}. Relaxed and unrelaxed groups with $M_{Halo} \leq 10^{13.7} M_\odot$ are considered; the cut in halo mass is implemented to minimize the physical cross-talk between $\delta_{LSS}$ and $M_{Halo}$ at high halo masses.  The  semi-analitic models  of Guo et al.\ are shown as a dotted black lines; in the bottom panels, they are also split in  bins of $R/R_{vir}$, as done for the ZENS data.  The black arrow indicates the direction of increase in density for the models, from sparse to dense environments. \label{f3a} }\end{center}
\end{figure}

\begin{figure} \begin{center}
\includegraphics[width=0.45\textwidth]{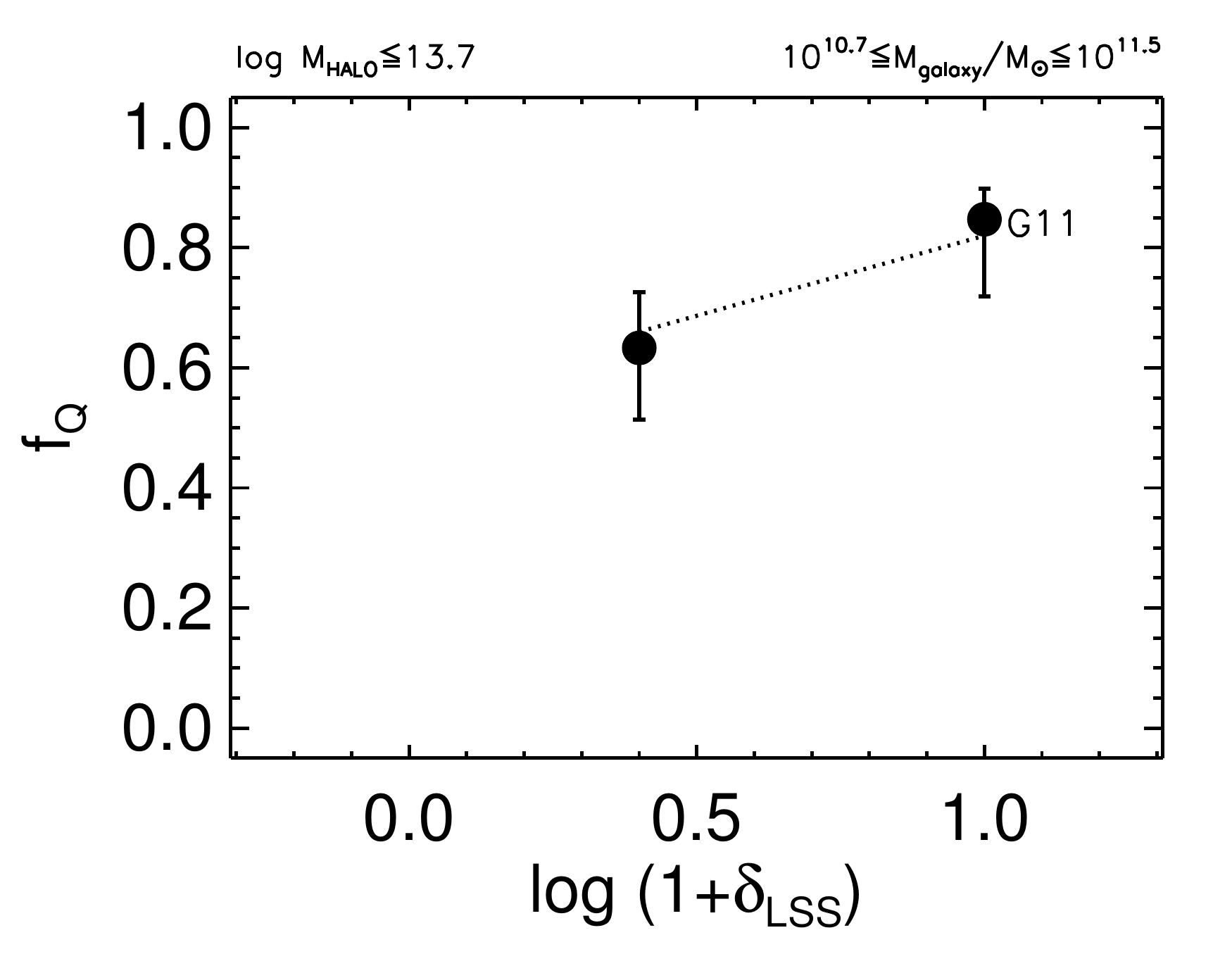}
\caption[]{ As for the top panel of  Figure \ref{f3a}, but now for the galaxy stellar mass bin $10^{10.7}-10^{11.5}   M_\odot$.     \label{f3b} }\end{center}
\end{figure}
\end{subfigures}

\section{Insights on quenching from morphologies}\label{discuss}

The main results from the previous Sections can be summarized in terms of two simple statements: $(i)$ the quenched fraction of satellites has a strong inverse dependence on halo-centric distance, but does not depend (much) on group halo mass and, especially for $M_{galaxy}\lta 10^{11} M_\odot$, also not on the surrounding LSS density; and $(ii)$  the morphological mix of the quenched satellites $f_{Q_{ETG}}$ does not appear to depend at all on radius, halo mass or LSS density, and therefore, by extension, is also independent of the quenched fraction $f_Q$.  We discuss these findings below in the context of the mass- and environment-quenching picture of Peng et al.\ (2010), and in comparison with some representative traditional SAMs.

\subsection{A similar impact of mass- and satellite-quenching on morphologies}\label{sjl}

Mass-quenching is the process that controls the overall mass-functions of star-forming and passive galaxies (Peng et al.\ 2010); it is clearly associated with the mass of the galaxy, although it is hard to distinguish whether it is the stellar mass, halo mass or even black hole mass that is relevant (see discussion and references in Section \ref{intro}). Environment-quenching appears to act only on satellites (e.g., van den Bosch et al.\ 2008, Peng et al.\ 2012, Wetzel et al.\ 2012, Kova\^c et al.\ 2013), and such satellite-quenching has been shown to be independent of stellar mass (when expressed as a differential quenching efficiency as in Peng et al.).  

Given the well-known correlation of early-type morphology with mass, a simple hypothesis would have been to suppose that mass-quenching produces early-type morphologies while satellite-quenching is not associated with any morphological transformation and adds quenched disk systems to the total population of passive galaxies.   Based on our results, we can rule out this simple idea. 

At a given stellar mass, some (constant) fraction of satellites should have been quenched by mass-quenching, independent of environment (Peng et al 2010).  We would expect this to also be the $f_Q$ of satellites at large radii $R > R_{vir}$.  
This is tantamount to saying that there is no satellite-quenching per se at these large radii. This is not inconsistent with  reports of alleged environmental quenching effects on galaxies at very large  distances from the centres of  massive clusters (e.g., an enhanced red fraction and reduced HI emission;  \citealt{2013arXiv1303.7231W} and references therein), since these effects are largely contributed by central  galaxies of lower-mass halos  and disappear when satellites-only are  considered (Wetzel et al.\ 2012).  Note also that, as shown in Appendix \ref{allrelaxed},  most of the massive satellites that are found at very large halo-centric distances are members of unrelaxed groups without a well defined center, which casts some doubts on their identification as $R>R_{vir}$ satellites of the given halo.  

Using the quenching efficiencies for mass-quenching of Peng et al.\  and taking into account a $\sim$0.2 dex offset between their stellar masses (actual, including return to ISM) and ours (integral of SFR), we would estimate  that of order 40\% and 70\% of satellites, respectively, in our low- and high stellar mass bins should have been mass-quenched.  This is consistent with what we observe at large radii in the low mass bin but, in the high mass bin, the observed $f_Q$ at $R > R_{vir}$ is apparently lower than this.  However, the number of galaxies involved is quite small (see Figure \ref{fA2}) and the corresponding statistical uncertainty quite large.   

For definiteness, we set the mass-quenched $f_Q$ value in the high mass bin to $f_Q = 0.6$, which is only about 1$\sigma$ discrepant from our observed value at $R > R_{vir}$ and is closer to the P10 expectation, but we stress that this choice is not critical to what follows.  The adopted mass-quenched $f_Q$ are shown as horizontal red lines in the top-right panels of Figures \ref{f1a} and \ref{f1b}.  

The observed increase in $f_Q$ with decreasing halo-centric radius in these Figures reflects the increased importance of satellite-quenching towards the centres of the haloes (we hatch the relevant areas in the top-right panels of Figures \ref{f1a} and \ref{f1b} to guide the eyes).    If mass-quenching and satellite-quenching had different morphological outcomes, then this would produce a clear trend with halo-centric distance in the morphological composition of the quenched satellite fractions shown in Figures \ref{f2a} and \ref{f2b}: the morphology associated with mass-quenching would dominate at large group radii, and that associated with satellite-quenching would become increasingly more important (up to about a 50:50 split) towards the central group regions where the $f_Q$ has almost doubled.  Put another way, since the radial variation of $f_{Q}$ is due to the variation in relative importance of the two quenching channels, we would expect, if these channels had different morphological signatures, to then see a variation of $f_{Q_{ETG}}$ with $f_Q$. 
We see no such variation in $f_{Q_{ETG}}$ and, as a result, we can reject this simplest hypothesis.   The obvious conclusion is either that neither of the quenching channels induces a morphological transformation, or that both of the two quenching channels result in the same morphological changes.   

To distinguish between these two possibilities, we can compare the morphological mix of the quenched satellites with that of the star-forming satellites at large radii, $R > R_{vir}$, since these are presumably representative of the progenitors of the quenched population.   The fraction of $R > R_{vir}$ star-forming satellites that are morphological early-types is substantially lower, in each mass bin, than the $f_{Q_{ETG}}$ that we have found for the quenched satellites, i.e. 15\% and 35\% for the low and high stellar mass bins as compared with 50\% and 65\%, respectively.  

This could be taken  to indicate that quenching is associated with {\it structural} changes in the galaxies, i.e.\, changes associated with the underlying mass distribution such as the growth of stellar mass in the bulge (or inner disk) components. It is   well established that quenched galaxies have radial surface brightness profiles that are steeper than those of star-forming galaxies and well described by high S\'ersic indices (e.g., Kormendy et al.\ 2009 and references therein), reflecting  high central mass densities  (e.g.\, Fang et al.\ 2013 and references therein). Also,  since the pioneering work of  Mihos \& Hernquist (1994),  inward flows of gaseous material have long been recognised to take place  during galaxy mergers and through  disk instabilities  (e.g., Friedli \& Benz 1995).  Such inward gas flows have been proposed to likely contribute to the growth of stellar mass in galactic bulges (see e.g.\, Immeli et al.\ 2004; Courteau et al.\ 1996;  MacArthur et al.\ 2003; our own work, i.e., Carollo et al.\ 1997, 1998, 2001, 2007 and Carollo 1999, 2004; the comprehensive review by Kormendy \& Kennicutt 2004 and references therein; Bournaud et al.\ 2011, and many others). This idea has been further developed in   more recent theoretical developments (e.g.\, Dekel \& Burkert 2013) and is supported by observational evidence at high redshifts (e.g.\, Genzel et al.\ 2006; see also Cameron et al.\ 2011) and, at least for low-mass bulges, also in the local Universe, where   small  bulges show signs of `rejuvenation' of their stellar populations, with up to 10-30\% of their total mass consistent with having formed in the past  few Gyr  (Thomas \& Davies 2006; Carollo et al.\ 2007). 
Since both mergers and disk instabilities may be directly or indirectly connected with galactic quenching, it is not implausible to draw a connection between quenching and the growth of bulges.  

Of course, establishing the direction of causality in any such change is very difficult.  Quenching could be directly linked to the growth of the bulge (e.g., Martig et al.\ 2013; Genzel et al.\ 2013), or the quenching process itself could be associated with bulge growth, or the prominence of the bulge could be linked to a third property (e.g. galactic mass or the presence of an AGN) that itself controls the quenching of star-formation.  Furthermore, changes in the disk surface brightness profiles following quenching are almost certain to occur once star-formation ceases, and this will inevitably increase the prominence of the bulges even if the bulge itself remains unchanged.

An alternative is therefore that  changes occur only  in the light distributions of galaxies i.e., in their light-defined  $B/T$ values and thus {\it morphologies}, due in particular to differential surface brightness fading after star formation ceases. 
There are theoretical suggestions whereby neither mass- nor environment-quenching mechanisms would produce a structural change. For example,  radio-mode AGN feedback can provide a quenching mechanism  that leaves unaffected the mass distributions of bulges and disks  (e.g., Gabor \& Dave' 2012).   Gas-removal once the galaxies become satellites will also preserve disks.   Since gas-removal as a satellite-quenching  mechanism  is implemented in several current traditional SAMs,  this gives us the opportunity to quantitatively compare our results on $f_Q$ and $f_{Q_{ETG}}$ with a couple of such SAMs in  Section \ref{mod}.

In the next two Sections we explore instead the  global, as well as  bulge and disk properties in the quenched and star-forming satellites populations, and   study the effects of  the fading of their disks (with no growth of bulges) to investigate whether structural as opposite to morphological changes are mainly responsible for the observed differences  between these populations.

\begin{subfigures}
\begin{figure} \begin{center}
\includegraphics[width=0.45\textwidth]{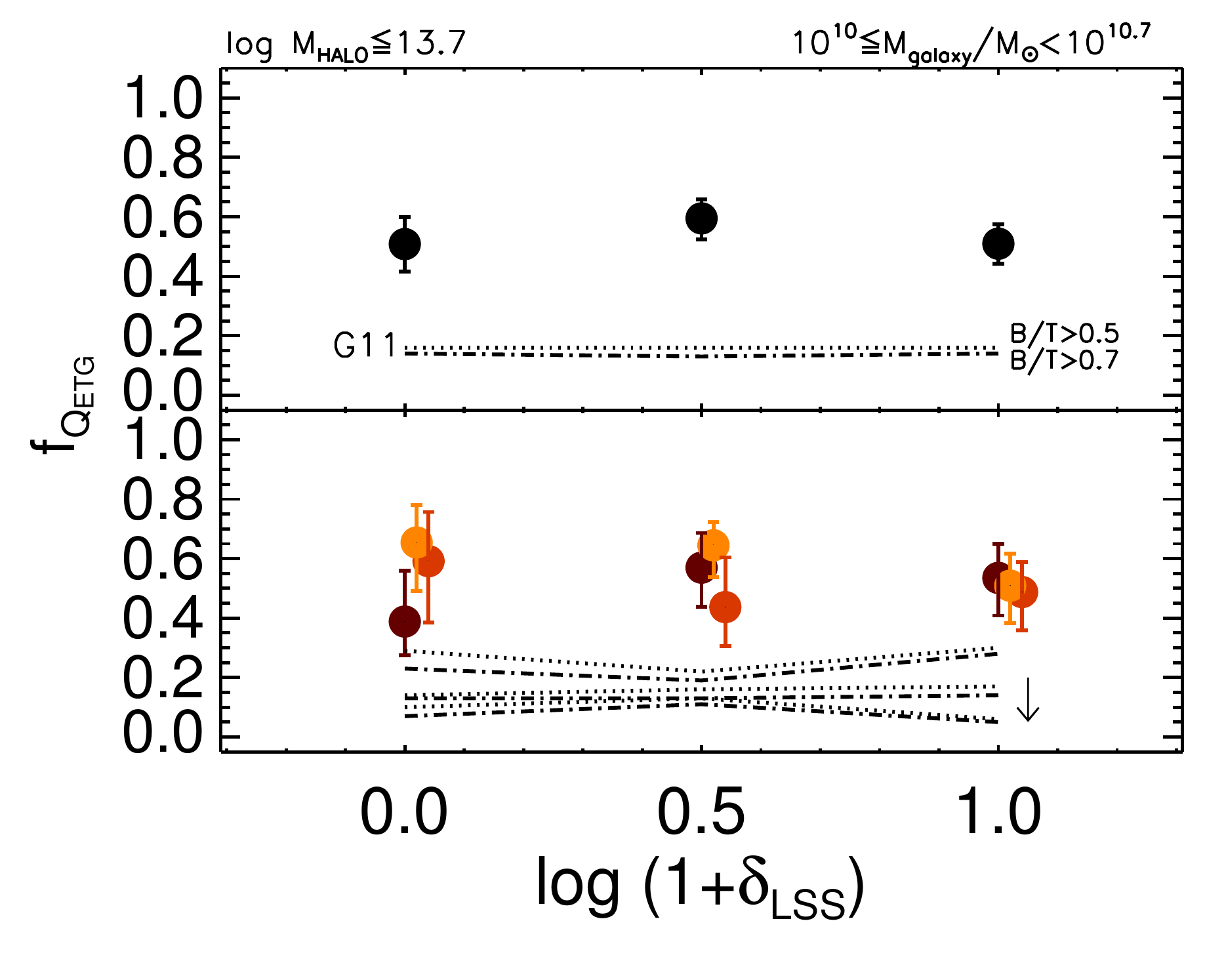}
\caption[]{  The fraction  of  quenched satellite galaxies {\it with an early-type morphology} as a function of  LSS density, for the galaxy stellar mass bin $10^{10}-10^{10.7}   M_\odot$.   Our LSS density parameter  reflects the location of a group in the cosmic web (not the local density of a given satellite within its group halo).  The top panel shows the total fraction integrated over halo-centric distance; the bottom panel shows the fractions in each of the bins of $R/R_{vir}$ highlighted  in Figure \ref{fA4}. Relaxed and unrelaxed groups with $M_{Halo} \leq 10^{13.7} M_\odot$ are considered.   The  semi-analitic models  of Guo et al.\ are shown as a black lines. Different line types indicate model early-type satellites defined as systems with a $B/T>0.7$ in stellar mass (dashed-dotted line); the dotted lines are for model satellite galaxies with  a stellar mass $B/T>0.5$.  In the bottom panels, the models are split in  bins of $R/R_{vir}$, as done for the ZENS data. The black arrow  indicates the direction of  increase in density for the models, from sparse to dense environments.\label{f4a} }\end{center}
\end{figure}

\begin{figure} \begin{center}
\includegraphics[width=0.45\textwidth]{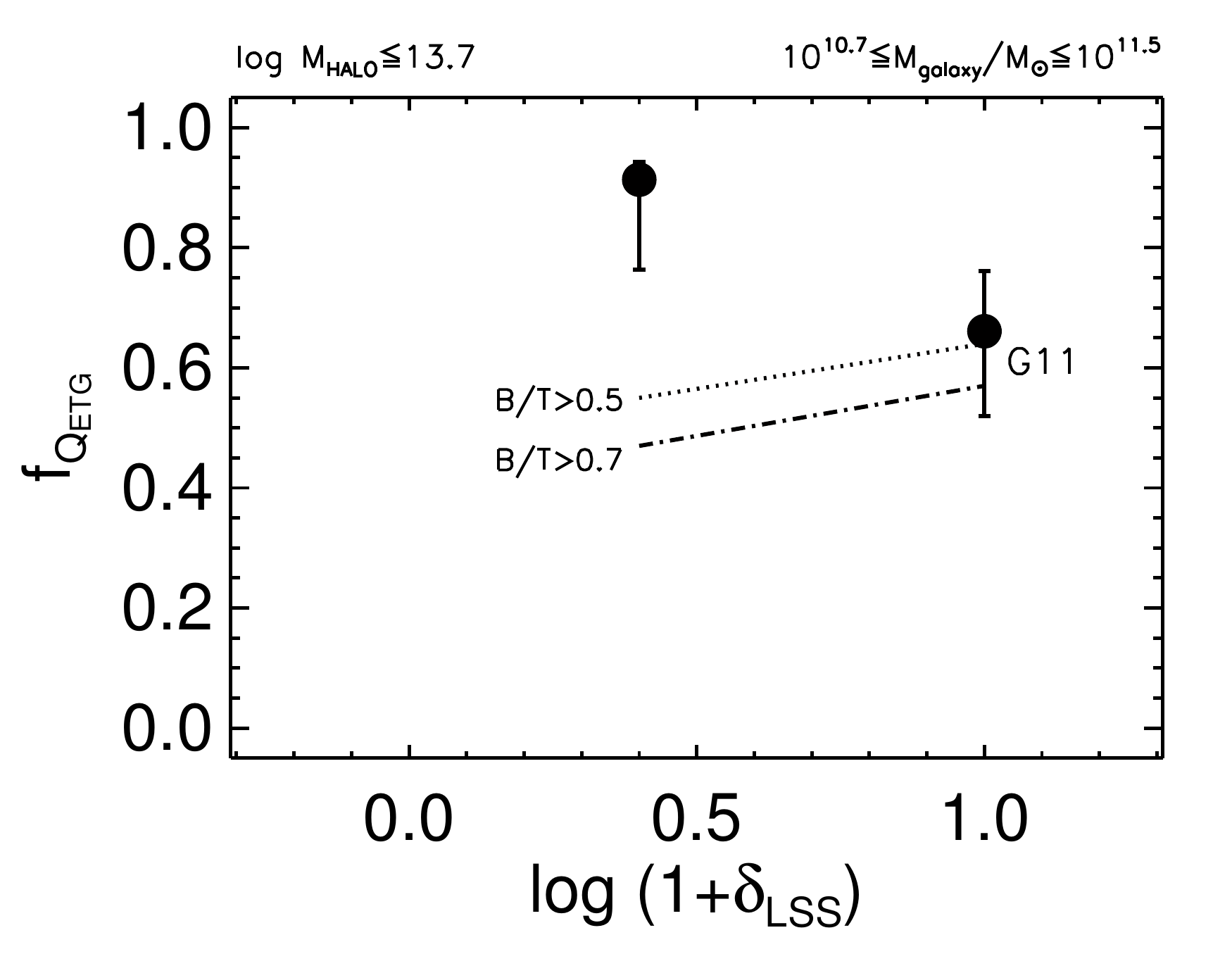}
\caption[]{ As the top panel of  Figure \ref{f4a},  this time for the galaxy stellar mass bin $10^{10.7}-10^{11.5}   M_\odot$.     \label{f4b} }\end{center}
\end{figure}
\end{subfigures}

\subsection{The global properties  of quenched and star-forming satellites and of their bulges and disks}\label{bulgedisk}

\begin{figure*}[htb]
\begin{center}
\includegraphics[width=0.8\textwidth,angle=0]{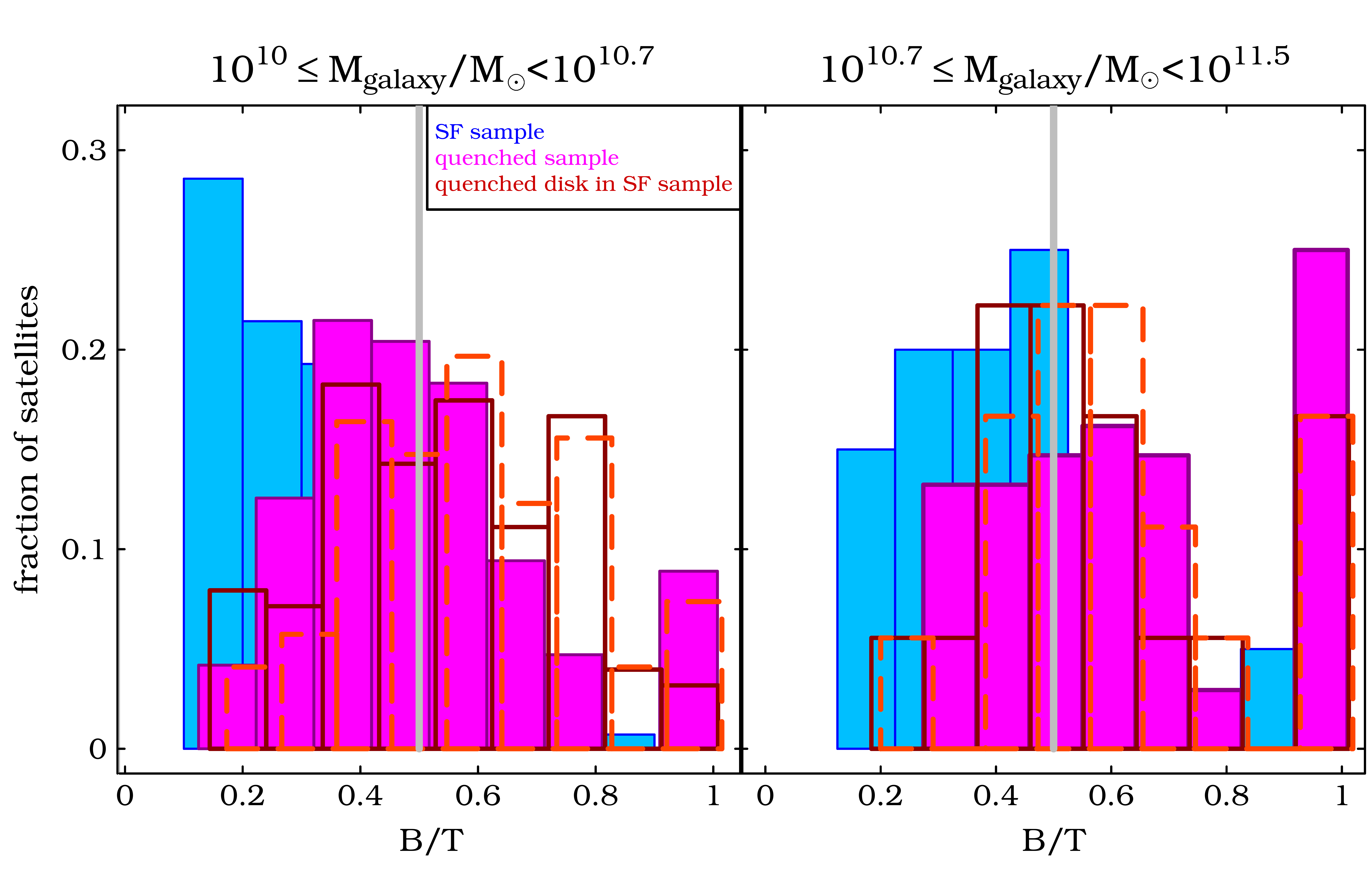}
\caption{\label{f5}Distributions of measured, light-defined $B/T$ ratios in the $I$-band for ZENS satellites in the two galaxy mass bins of our study. The magenta histograms correspond  to the samples of 
 quenched satellites that we study in this paper. Note that elliptical morphologies  are placed at a value of B/T=1.  The blue histograms show the distribution of $B/T$ for  the original sample of 
 ZENS  star-forming  satellites  within the same bins of  stellar mass.
The red histograms illustrate  the effect of  disk-quenching  on these star-forming satellites populations (dark red, solid = 1 Gyr passive evolution;  light red, dashed = 3 Gyr passive evolution). The vertical grey lines in both mass bins show  the $B/T$ threshold which we use   to separate  the two morphological classes of disk-dominated and early-type galaxies.}
\end{center}
\end{figure*}

We have argued in the previous section that the morphological outcome of the different quenching channels appears to be the same, even though these quenched satellites have different morphologies to the star-forming satellites, presumed to be representative of the progenitors of the quenched galaxies.
Figure  \ref{f5} shows in magenta the distribution of the measured light-defined $B/T$ ratios in our original sample of quenched satellites\footnote{The distribution of measured light-defined $B/T$ ratios for the original sample  of quenched satellites is here averaged over all environmental bins, and, given the independence of $f_{Q_{ETG}}$ on any environment, is very similar to the individual values plotted in Figures \ref{f2a} and \ref{f2b}.} and in blue that for the 
star-forming satellites with a measurable light-defined $B/T\ge0.1$. This represents virtually all (i.e., 99\% of) star-forming satellites in the high mass bin, and more than 75\% in the lower mass bin.   We exclude the $B/T < 0.1$ satellites because we are interested in quantifying the emergence of the bulges when the disk fades.  The question why there are very few bulge-less quenched galaxies at low masses remains open. As we discuss below, it may be that there is bulge growth associated with quenching in these low bulge systems, which may shift the progenitor pure-disk satellites into the $B/T\ge0.1$ sample after quenching, or alternatively a small $B/T\sim0.1$ bulge may be hidden by a bright star-forming disk and emerge and be measurable above the $B/T=0.1$ threshold when the disk fades after quenching. Note also that, to reduce statistical errors,  star-forming satellite at all halo-centric distances are included, but the following results do not change   when using only  the $R>R_{vir}$ star-forming sample.  
The median $B/T$ values for the  final star-forming samples 
are 0.30 and 0.43 in the low and high stellar mass bins; the corresponding values for the quenched satellites  are 0.47 and 0.60, respectively.  

\begin{figure*}[htb]
\begin{center}
\includegraphics[width=0.8\textwidth]{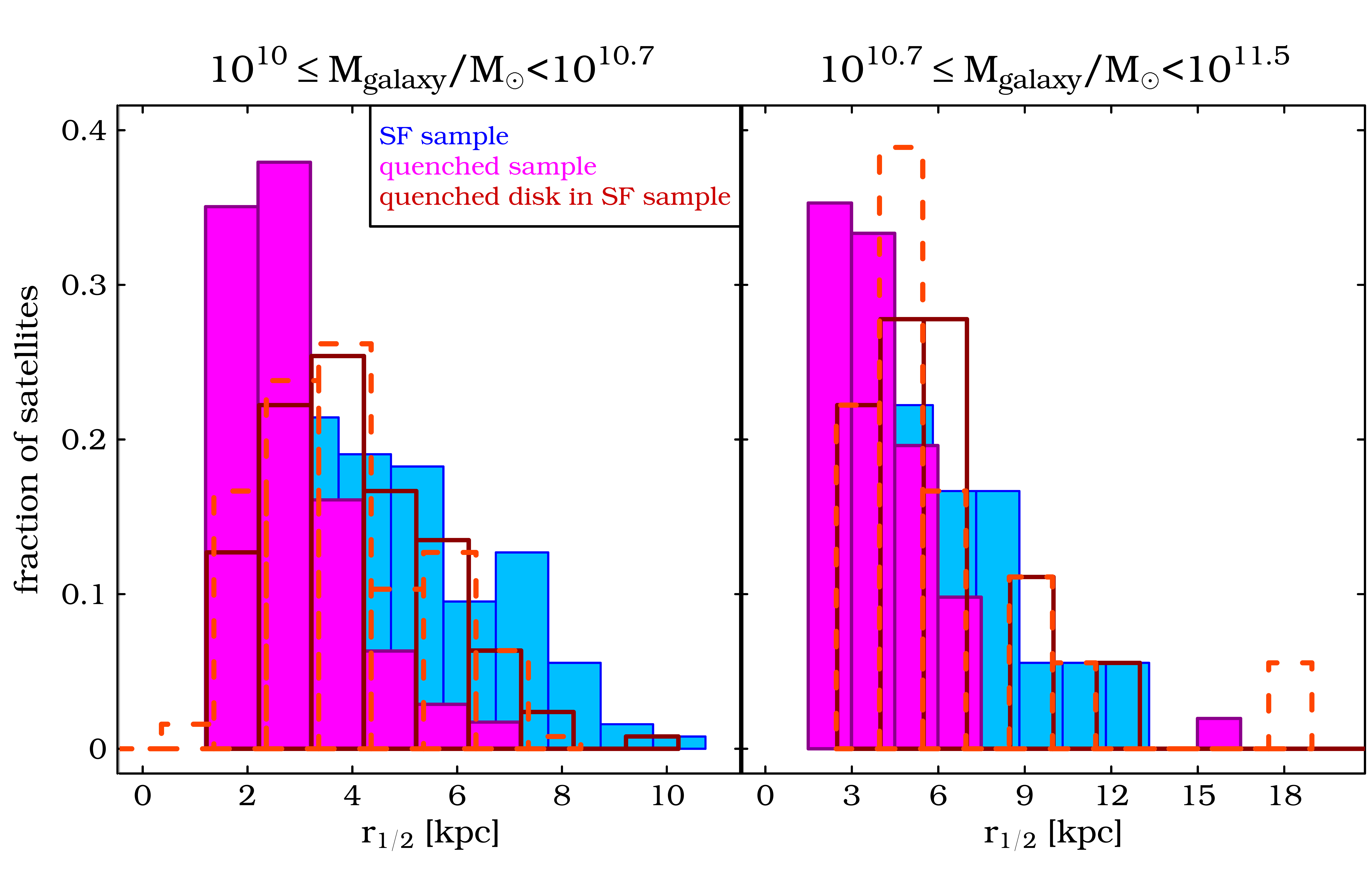}
\caption{\label{f6}Distributions of galaxy half-light radii in the $I$-band for ZENS satellites in the two galaxy mass bins of our study. In magenta we show the sample of 
 quenched satellites that we study in this paper.  In blue are the histograms  for  the original sample of 
 ZENS  star-forming  satellites  within the same bins of  stellar mass.
The red histograms illustrate  the artificially disk-faded satellites populations (dark red, solid = 1 Gyr passive evolution;  light red, dashed = 3 Gyr passive evolution).}
\end{center}
\end{figure*}

The quenched satellites also have smaller half-light radii  than star-forming satellites.  Figure \ref{f6} plots  the histograms of total galaxy half-light radii $r_{1/2}$ for the star-forming  (blue) and quenched (magenta) satellite samples: the average sizes of the quenched satellites are about 2.1kpc and 2.7kpc at low and high stellar masses, compared with 
3.4kpc and 6kpc for the star-forming samples.    These values reflect the global mass-radius relations for quenched and star-forming nearby galaxies (see e.g., the extensive discussion on Paper II and references therein).

The increased $B/T$ and the smaller half-light radii in the observed quenched satellites relative to the star-forming ones
could be taken to indicate some growth of stellar mass in the bulges of star-forming satellites when they are moved  from the star-forming `main sequence' (on the galaxy SFR vs.\ $M_{galaxy}$ plane, e.g., Daddi et al.\  2007) to the quenched population (either before quenching, i.e.,    in the gas flows regime, or after quenching, i.e.,  in the secular stellar disk bar-like instability regime, see Kormendy \& Kennicutt 2004 and references therein).
In order to study this further, Figure \ref{f7} shows the surface brightness profiles for the bulge and disk components  of the quenched and star-forming satellites. The broad bands show the  range of profiles which are then averaged (in log surface brightness) to produce an average profile, shown as a solid line.  The average star-forming disk profile is shown in blue, and  the average bulge in the star-forming satellites is shown in black. 
In  magenta and grey we plot the observed disk and bulge profiles in the quenched satellites.  

It is noticeable how similar the bulge profiles of star-forming and quenched satellites are at constant galaxy stellar mass, despite the substantially lower average light-based $B/T$ value measured for the former relative to the latter (although we discuss below a small effect in the inner profiles within 1 kpc).  In contrast, the surface brightness profiles of the disks in the quenched satellites appear to be overall fainter and steeper (i.e. with smaller exponential scale lengths) than those of the presumed-progenitor star-forming disks of the same integrated galactic mass.  

The fact that the overall bulge surface brightness profiles  are so similar in the quenched and in the star-forming satellites argues in principle against substantial structural changes in the bulge components due to quenching. The two most stable parameters for the bulges are the total luminosities and the central surface brightnesses (see Paper II).  This is also confirmed by the simulations that we described below in Section \ref{minimal}.  

We therefore plot in Figure \ref{f8}   the $I$-band bulge surface brightness within 1 kpc  versus the total bulge $I$-band luminosity for the quenched and star-forming satellites in the two bins of galaxy mass.   The median of the $I$-band bulge luminosities are very similar in the star-forming and quenched satellites, i.e.,  $I$-band apparent magnitudes of 17.01 versus 16.99 mag respectively in the low mass bin and 15.83 versus 15.88 at the high masses.  
There is a small shift  in the average bulge surface brightness within 1 kpc at constant $I$-band magnitude between the two satellite samples. This is better shown in the bottom panels of Figure \ref{f8} where we plot the residuals  $\Delta\langle\mu_{1kpc}\rangle$ for the bulges in the quenched (magenta) and star-forming (blue) samples, relative to the average relation between bulge $\langle\mu\rangle_{1kpb}$ and bulge $I$-band magnitude indicated by the dashed black lines in the top panels of Figure \ref{f8}. The linear parameters of these average fits  in the low and high galaxy mass bins are respectively given by: $a=1.05\pm0.05$,  $b=-0.68\pm0.2$ and $a=0.85\pm0.05$, $b=2.94\pm0.8$.

The small shift of order 0.3 mag in bulge $\Delta\langle\mu_{1kpc}\rangle$  is consistent with a small amount of stellar mass being added to the centres of the bulges of quenched satellites.  It is clear however that any such increase in the overall mass of the bulge cannot be substantial and in particular cannot significantly change the overall bulge ($I$-band) luminosity.

The comparison  between the disk profiles of star-forming and quenched satellites, on the other hand,  indicates that the change in the global light-defined $B/T$ and the reduction in the galaxy half-light radii between star-forming and quenched satellites 
are primarily associated with differences in their  disks.  Figure \ref{f9} shows the measured disk parameters for individual star-forming (blue) and quenched (magenta) satellites.  These indicate that the central disk surface brightnesses averaged within the innermost 1kpc do not change  very much (${\langle\mu\rangle}_{SF,1kpc}= 19.85$ [19.79]  mag vs.\ ${\langle \mu\rangle}_{Q,1kpc} =19.75$ [19.59] mag at low [high] galaxy masses), whereas the distribution of scale lengths of the disks in the quenched satellites is shifted to smaller values as compared with the scale lengths of star-forming satellites ($h_{SF}=$ 3.4[4.4] kpc   vs.\ $h_Q =$ 2.2[3.3] kpc again at low [high] masses).

In this context it is important to  note that, since star-formation is mostly associated with disks, at least at late epochs (and possibly always), a morphological change in the direction of an increase of the light-defined $B/T$ would be expected as an aftermath of quenching, even in the absence of structural (i.e., mass-based B/T) changes, simply due to the fading in surface brightness of a star-forming disk once star-formation ceases.   A fading of the disks would increase the apparent $B/T$ and shift galaxies towards earlier morphological types, potentially explaining the observed $B/T$ of the quenched satellites. Fading of a star-forming disk in the aftermath of quenching clearly also reduces the overall galaxy half-light radius of a galaxy. This would explain also the smaller sizes of quenched satellites, and has previously been invoked to help reconciling the differences between  predicted and observed sizes functions of $\sim M*$ quenched galaxies at high redshifts in the context of the apparent change in size of the passive galaxy population \citep{2013ApJ...773..112C}. We therefore explore  below  in more detail the quantitative impact  of disk fading on the global structure of galaxies.

\subsection{Morphological changes associated with the fading of disks in the aftermath of disk-quenching}\label{minimal}

To  quantify the impact of post-quenching disk-fading on the $B/T$ and sizes of galaxies, a first approach is simply to take the bulge-to-total ratio $B/T$ of the star-forming disks, i.e.\ $({B/T})_{SF}$, and recompute the new `quenched' $(B/T)_Q$  that would be obtained with a reduced disk contribution.  The resulting quenched $({B/T})_Q$ will be a non-linear function of the initial $({B/T})_{SF}$ and the reduced surface brightness of the quenched disk, which we represent by $\xi = D_Q/D_{SF} < 1$. Specifically, with:

\begin{equation}
(B/T)_{SF}={{B} \over {B+D_{SF}}},
\end{equation}

and under the assumption that disk fading is not accompanied by any change in the bulge component, $B$, we get:

\begin{equation}
(B/T)_Q =  {{B} \over {B+D_Q}} = {(B/T)_{SF} \over \xi + (1-\xi)(B/T)_{SF}}.
\end{equation}

For a disk fading of about 1 magnitude, i.e. $\xi \sim 0.4$, a typical star-forming galaxy with an original $({B/T})_{SF} = 0.4$ will have $({B/T})_{Q} = 0.63$ after quenching, sufficient to make it an early-type galaxy.

\begin{figure*}[htb]
\begin{center}
\includegraphics[width=0.8\textwidth,angle=0]{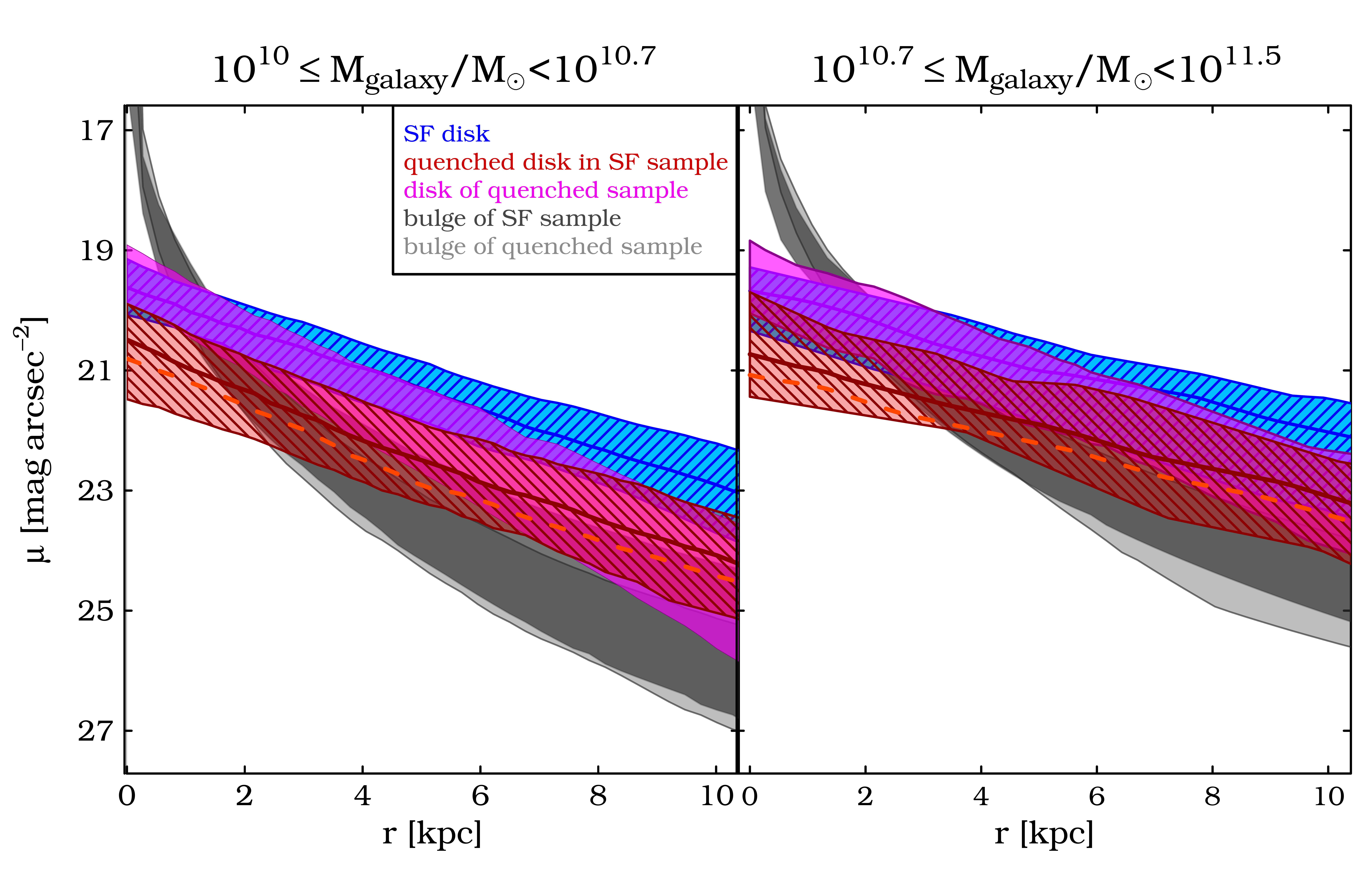}
\caption{\label{f7} Average $I-$band surface brightness profiles for the bulges and disks in the sample of ZENS satellites 
 in  the two mass bins of our study.
The solid line show the median surface-brightness profiles, the shaded areas indicate the 25th and 75th percentiles of the distributions.
The bulges and disks of the  quenched  satellites are shown with the light-grey and magenta curves, respectively.
The   corresponding components for the star-forming  satellites are shown in dark-grey and   blue, respectively. In red we show the artificially faded disks (dark solid and light dashed lines for the 1 and 3 Gyr passive evolution models, respectively).  For clarity, the scatter for the disk-fading models is shown only for the 1 Gyr passive evolution  simulations (and is similar for the 3 Gyr passive evolution model).}
\end{center}
\end{figure*}

\begin{figure*}[htb]
\begin{center}
\includegraphics[width=0.8\textwidth]{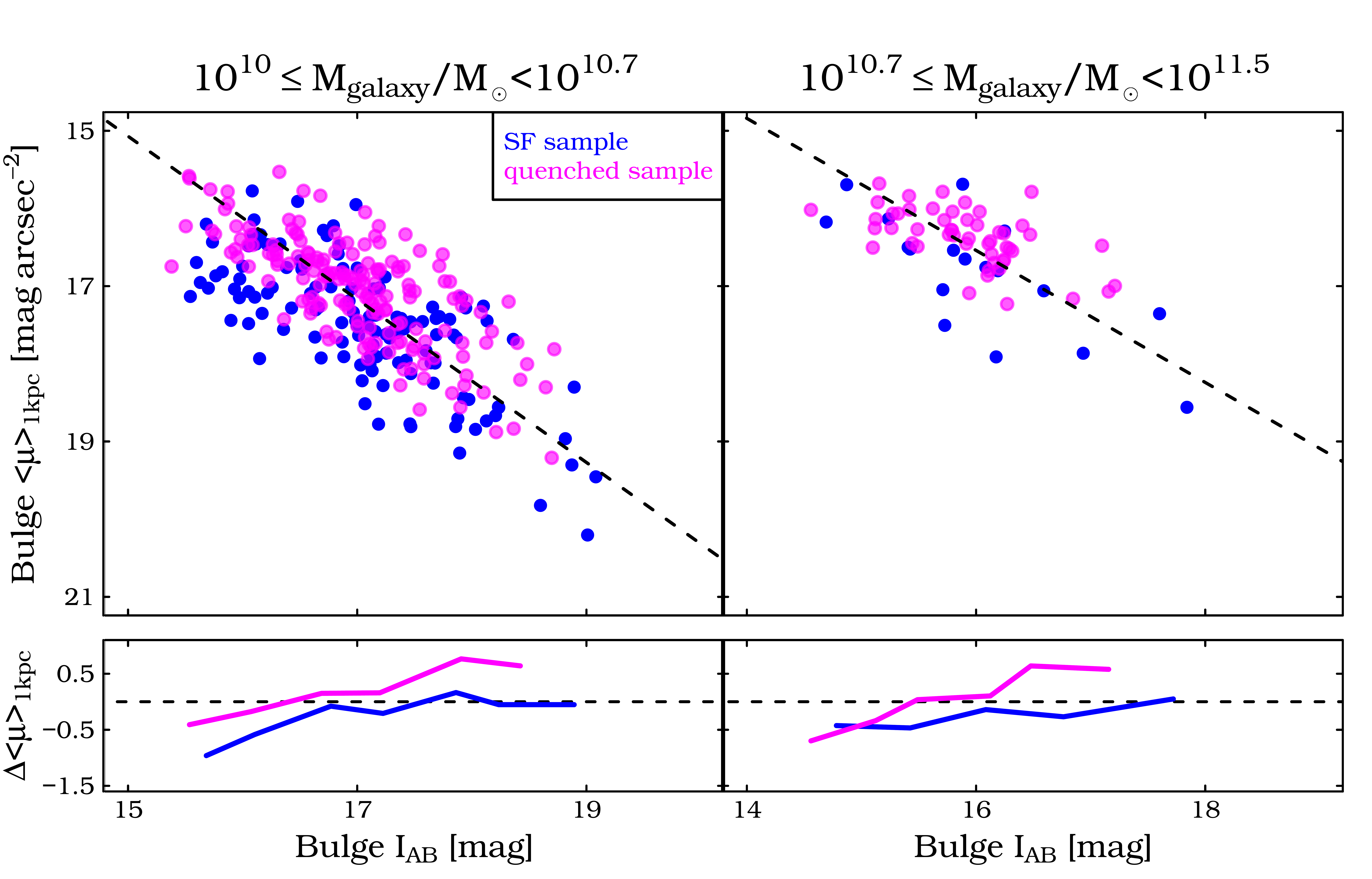}
\caption{\label{f8} {\it Top panels:} The central bulge surface brightness within 1 kpc plotted against  the total $I$-band bulge luminosity in the low (left) and high (right) galaxy mass bins. As in the previous Figures, the star-forming sample is plotted in blue  and the quenched sample in magenta. The black dashed line shows the average relation  for the total sample of quenched plus star-forming satellites. {\it Bottom panels:} The corresponding residuals relative to the average (black dashed lines) relations  above  for the quenched (magenta) and star-forming (blue) satellites. }
\end{center}
\end{figure*}

\begin{figure*}[htb]
\begin{center}
\includegraphics[width=0.8\textwidth]{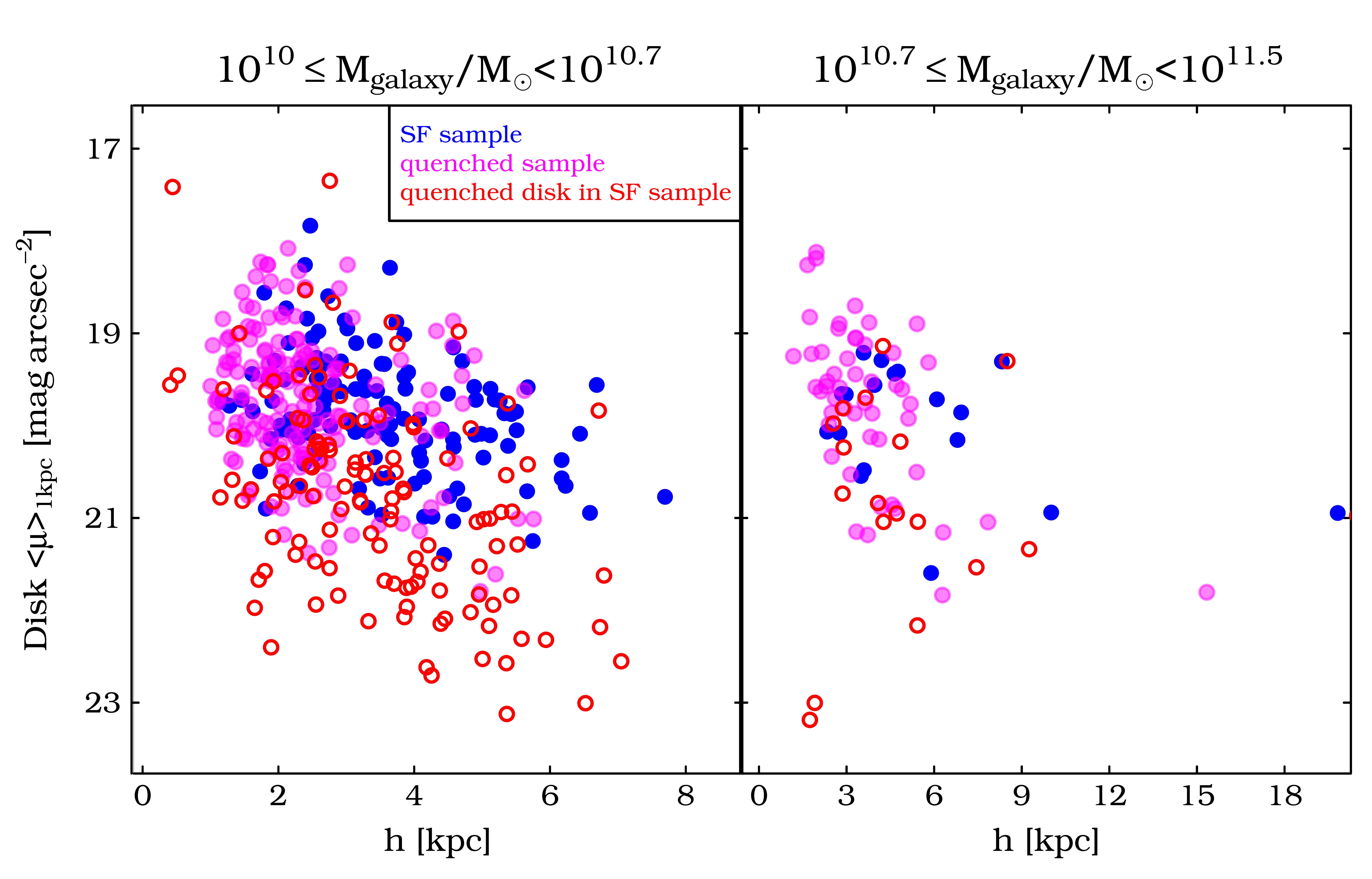}
\caption{\label{f9} The central disk surface brightness within 1 kpc plotted against  the disk scale length. The blue filled circles are the observed disks for the star-forming satellites, the red empty circles  show the 1 Gyr passivized disks, and the magenta dots are the observed disks in the quenched satellites. }
\end{center}
\end{figure*}

We can look at this from a different perspective by looking at the actual surface brightness profiles of individual star-forming galaxies in ZENS as decomposed into their bulge and disk components.  Specifically, we use the light-defined $B/T$ decomposition for each star-forming galaxy that we presented in Paper II and show in Figure \ref{f7},  and we generate a simulated $I$-band image with the same bulge and disk structural parameters, except that the disk surface brightness is reduced, uniformly at all radii, by a certain amount $\Delta\mu_{fade} = -2.5 log \xi$.  We consider fading of 1.0 to 1.5 magnitudes, corresponding to the fading of a disk that has been forming stars at a constant rate prior to being quenched 1 to 3 Gyr before the current epoch of observation (using  for this estimates the Bruzual \& Charlot 2003 stellar population synthesis models with a \citealt{Chabrier_2003} Initial Mass function and a Solar metallicity). These artificially-quenched images are then convolved with the typical ZENS $I$-band Point Spread Function and an appropriate level of noise is added. We then carry out the same bulge+disk decomposition fits as done for the real ZENS galaxies (see Section \ref{morphs} and Paper II for details).   The results\footnote{As mentioned above, we also used these simulations to test which bulge and disk parameters can be considered as reliable. As also discussed in Paper II, the total magnitudes,   scale lengths and central surface brightnesses (within 1 kph) of the disks, as well as the bulge total magnitudes and central surface brightnesses, are well recovered even after disk fading. In contrast, the bulge S\'ersic indices and half-light radii are highly degenerate and unstable. We thus limit the comparisons  between bulges and disks of different satellite samples to only those parameters which can be reliably measured.} are shown in red in Figures \ref{f5}, \ref{f6}, \ref{f7}  and  \ref{f9}.

It is clear from Figures \ref{f5} and  \ref{f7}   that fading of the disk enhances the prominence of the central bulge leading to a higher light-defined $B/T$.  The light-defined $B/T$ values for the  average galaxy profiles change at $10^{10}<\Msol<10^{10.7}\Msol$ from 0.24 for the star-forming average profile, to 0.44 or 0.52 for the two fading models being examined; the corresponding change  at  $10^{10}<\Msol<10^{10.7}\Msol$ is from 0.36 to 0.55 or 0.63.   The median $B/T$ values of the distributions of individual $B/T$ shown in Figure \ref{f5}
change from 0.30 and 0.43 for the original star-forming galaxies  
  in the low and high stellar mass bins,  to  an average of 0.50 and 0.57 for both low and high masses for the 1 Gyr (dark red solid) and 3 Gyr (light red dashed) disk-fading models, respectively. 
These values are quite close to the 0.47 and 0.60 medians  of the $B/T$ distributions of real quenched satellites  in the two mass bins.  In both stellar mass bins  a Kolmogorov-Smirnov test indicates that the measured $B/T$ distributions of the quenched satellites and the artificially faded star-forming satellites are fully consistent with each other (with a slightly better agreement for the 1 Gyr disk-fading model). 

Figure \ref{f6} shows however that the galaxy half-light radii $r_{1/2}$ obtained from these uniform fading models (e.g., 3.2 kpc and 4.8 kpc in the two mass bins using the representative 1
Gyr post-quenching model) remain  still significantly larger than those the actual quenched satellites, 2.1 kpc  and 2.7 kpc respectively. The disk scale lengths of the artificially faded disks also are too large (3.28 kpc and  4.2 kpc at low and high masses, respectively),  and their central surface brightnesses too faint (20.77  and 20.89 mag; see Figure \ref{f9}).

A reduction in scale length would however be expected due to a differential fading of the disks with galactic radius, e.g. due to a stellar age gradient within the disks. Smaller sizes could also arise if disks were smaller, at given mass, at earlier epochs, i.e. when the  quenched satellites were actually quenched.  Both would be a natural consequence of an inside-out growth of individual disks.  We can quantify the first effect (differential fading) as follows.  We assume that both the original (star-forming) and the subsequently quenched disks can be represented by exponential light profiles with respective scale-lengths $h_{SF}$ and $h_{Q}$ and central surface brightnesses $\mu_{SF,0}$ and $\mu_{Q,0}$. The radial change in surface brightness will itself be an exponential with scale length: 

\begin{equation}
h_{fade} = {{h_{SF} h_Q}\over{h_{SF}-h_Q}} = (h_{Q}^{-1}-h_{SF}^{-1})^{-1}. 
\end{equation}

With: 

\begin{equation}
\mu_{SF}(r) = \mu_{SF,0} + 1.085 r h_{SF}^{-1}
\end{equation}

 and
\begin{equation}
\mu_{Q}(r) = \mu_{Q,0} + 1.085 r h_{Q}^{-1},
\end{equation}

the fading $\Delta\mu(r)_{fade}$ at different radii will be given by:

\begin{equation}
\Delta\mu(r)_{fade} = (\mu_{Q,0} - \mu_{SF,0}) + 1.085 r (h_{Q}^{-1}-h_{SF}^{-1}).
\end{equation}

An apparent halving of the scale length, i.e., $h_Q \sim 0.5 h_{SF}$ (roughly  what we see in our data) corresponds therefore to the application of a differential fading of the same scale length as the original disk, $h_{fade}\sim h_{SF}$, and to $\Delta\mu(r)_{fade}=1.085 r h_{SF}^{-1}$. 
Such little fading at the center, i.e., $\mu_{Q,0}-\mu_{SF,0}\sim0$ and    1 magnitude of fading at the original scale length of the star-forming disk (consistent with the 1 Gyr passive evolution model above) can thus completely explain the smaller sizes of quenched satellites relative to disk-faded star-forming satellites.  

The overall conclusion from these analyses is that simple fading of the disks after star-formation ceases can, in principle, explain most of the change in the observed (light-defined) $B/T$ and in the mean half-light radii, that are seen relative to a plausible set of star-forming progenitors, without the need for any substantial mass growth or other changes in the bulge components. This therefore also supports the idea that neither mass-quenching nor satellite-quenching produces  a significant structural change in the stellar mass distribution of satellite galaxies.   Clearly, this conclusion is based on analysis of galaxies at the present epoch, albeit reflecting quenching that happened at earlier epochs.   It therefore may or may not hold for quenching occurring at very much earlier times, e.g at $z \sim 2$ when passive galaxies first appear in substantial numbers (e.g. Ilbert et al 2013).

\subsection{SAM predictions  for the environmental dependence of the  quenched satellite fraction    and  its morphological composition}\label{mod}

Generally speaking, a long-standing problem of SAMs  remains the over-prediction of the total number of quenched satellites \citep{Bower:2006bh,De-Lucia:2006zr,Font:2008uq,Kimm:2009ij}.  Even recent models which  implement  more accurate prescriptions for  the removal of the  satellites' gas reservoirs still predict too many galaxies in the green valley between the red-sequence and the blue-cloud  \citep{Font:2008uq}. This is evident especially in Figure  \ref{f1a}, in the low galaxy mass bin of our ZENS analysis and, although in a much reduced form, also  in Figure  \ref{f1b} in our high galaxy mass bin. In both Figures, the dotted and dashed black lines show respectively, for the  SAMs of  \citealt{De-Lucia:2006zr}(DL06) and \citealt{Guo:2011ve} (G11)\footnote{Galaxy catalogs made available online at:  \\ \phantom{aaaaa}http://gavo.mpa-garching.mpg.de/Millennium/}, the dependence of $f_Q$on the group halo mass (left panels) and halo-centric distance (right panels).  

The main improvements in the G11 model relative to the earlier DL06 prescription are  a more accurate calculation of the ram pressure from the intra-group medium, and a more gradual resulting stripping of gas from satellites as these move towards the inner group regions.  While the gradual gas stripping narrows the gap between data and theoretical predictions by lowering the quenched satellite fractions in the G11 relative to the DL06 SAMs, the G11 model still over-predicts  the absolute number of quenched satellites, in particular  in our low galaxy mass bin.  Most likely both the star formation histories of  low mass galaxies at M$_{galaxy}<M*$ must be  improved and, at the same time, these galaxies should be prevented from loosing their gas too easily. 

A number of successful matches between data and models   are however similarly important  to highlight.  The recipes implemented in both the  DL06  and G11 models do reasonably well in predicting almost no dependence of the total (i.e., integrated over $R$) quenched satellite fraction $f_Q$ on group halo mass and, as discussed above, the dependence of $f_Q$ on $\delta_{LSS}$, both results in agreement with our data. Also, both models  predict  some  decrease in the quenched satellite fraction with increasing halo-centric distance, as seen at all galaxy masses in our  sample. 

In the low galaxy mass bin, although the instantaneous gas stripping algorithm at work in the DL06 models leads to a somewhat milder decrease in $f_Q$ with mass than the gradual gas stripping algorithm in G11 (which gives satellites the time  to establish a more significant halo-centric radial trend in the total quenched fraction), the amplitude of the radial decrease in $f_Q$ in our data is roughy consistent with both models. We note that, in G11, the timescales for establishing the radial trend of $f_Q$ with $R/R_{vir}$ are typically $\sim4$ Gyrs, i.e.,  a factor $\sim2$ longer than the  estimates for  $\tau_{quench}$ that are given by  our analysis of the color properties of satellites in our Paper III, and by several independent studies (Balogh et al.\ 2000; Wang et al.\ 2007; Weinmann et al.\ 2009; Feldmann et al.\ 2010, 2011; Rasmussen et al.\ 2012).   Together with the large over-prediction of quenched  galaxies with a disk morphology (that we discuss below), this suggests that gas-removal from disk satellites orbiting within  a common group halo is likely a faster but less universal process than what  is currently implemented in the SAMs. This is also suggested by some numerical simulations of satellite evolution in group potentials (e.g., Feldmann et al.\ 2010).

In the high galaxy mass bin, the G11 model reproduces the observed trend of $f_Q$ with $R/R_{vir}$ at distances $R<R_{vir}$, although it shows a rather flat relation at halo-centric distances $R>R_{vir}$; the DL06 model fails instead to predict the observed decrease of $f_Q$ with $R/R_{vir}$ even within the virial radius.   Given this somewhat better performance of G11 in reproducing all observed global trends of $f_Q$ in both stellar mass bins, we adopt this model as the fiducial one for the comparison of the observed trends of $f_{Q_{ETG}}$ with $M_{Halo}$ and $R$ (and $\delta_{LSS}$).     

G11 define as morphological  early-types galaxies with a bulge-to-total ratio $B/T > 0.7$ in stellar {\it mass}; in ZENS we adopted a similar $B/T$ threshold to select our early-type sample, but measured in $I$-band {\it light}. In Figures \ref{f2a},b  and \ref{f4a},b we overplot the model predictions for the $B/T>0.7$ sample as dashed-dotted black lines. Given the difference in the definitions between data and models, we also plot equivalent model lines for a morphological mass selection equal to $B/T>0.5$, so as to explore the effects of the precise $B/T$ threshold  value. The models are then treated as were the observations: e.g., in the low galaxy mass bin (Figures \ref{f2a} and \ref{f4a}), where the number of galaxies enables a  detailed investigation, the models are shown both integrated over the second environmental parameter  (halo-centric distance for the plots with halo mass as the abscissa, and halo mass for the plots with  halo-centric distance as the abscissa; top panels of the Figures) and also split in the same bins of the second environmental parameter as done for the data (bottom panels). 

The G11 model well reproduces the observed constancy of the morphological mix with any environment ($M_{Halo}$, $R$ and $\delta_{LSS}$) at all galaxy mass scales of our analysis. 
In the high galaxy mass bin the model only slightly underpredicts the high $f_{Q_{ETG}}$ (see Figures \ref{f2b} and \ref{f4b}),  but
in the  low galaxy mass bin the amplitude of $f_{Q_{ETG}}$ is underpredicted by a large factor of $\sim$2-3 (see Figures \ref{f2a} and \ref{f7}). This reflects the steep decrease  of  the total fraction of early-type galaxies in G11 when crossing the $M_{galaxy}  \sim 10^{11} M_\odot$  threshold from higher to lower masses (red line in Figure 4 of G11).

At $M_{galaxy} \gta  10^{11} M_\odot$, the main mechanism for bulge formation in the G11 model  is through galaxy-galaxy mergers that either occurred  at early times, or involve  gas-poor collisions of  bulge-dominated, already-quenched galaxies. These conditions lead to the relative high fraction of quenched early-type satellites at these high galaxy masses, and its independence  on the present-time  environment.  

In contrast, at lower masses, the dominant mechanism to form spheroids in the G11 model is via disk instabilities. These instabilities depend on internal galaxy properties only, and not on (any) environment (see Equation 34 in  G11).  Furthermore, the gradual environmental `strangulation'  that is implemented  in the G11 model  leads to satellite-quenching by the stripping of the hot gas halos while leaving  their pre-quenching morphologies unaffected\footnote{Note that there is also tidal disruption of the stellar structures included in the G11 model,  but when the condition for it to occur is satisfied, the whole galaxy is suddenly disrupted, which leads to it dropping out from the satellite sample.}. This leaves the morphological mix of the quenched satellite fraction equal to that of the progenitor star-forming satellite population -- which, as commented above, does not depend  on any environment, as indeed seen in our data.  

The   conclusion that we draw by  comparing  our data with the above SAMs, and in particular with the G11 model, is that in the regime of galaxy masses where environmentally-driven satellite quenching starts being relevant, i.e., at  galaxy masses below $M*$, the models $(i)$ overpredict the total quenched satellite fraction in all environments; $(ii)$ under predict the fraction of satellites with an early-type morphology in all environments, i.e., effectively they over predict the fraction of quenched satellites which retain a substantial disk component, and, at the same time, $(iii)$ roughly reproduce the slopes (but not  the amplitudes) of the relations of both $f_Q$ and $f_{Q_{ETG}}$ with  $M_{Halo}$, $R/R_{vir}$ and $\delta_{LSS}$.  We note that, in the model, these slopes  are achieved with a physical prescription  for environment-quenching (via hot-halo gas stripping) that, as discussed above, does not alter the morphologies of  the progenitor star-forming satellites, as we also argue in Section \ref{sjl}. On the other hand, the severe model-data discrepancies highlighted in $(i)$ and $(ii)$ together indicate that the model subjects too many star-forming disk satellites to environmental-quenching, thereby leading, at least at these galaxy mass scales, to too many quenched satellites with a disk-dominated morphology.

\section{Summary and concluding remarks}\label{con}

Using the ZENS sample of nearby group halos in the mass range $M_{Halo}\sim10^{12.2-14.9} M_\odot$, we have   explored how, in the local Universe, the fraction $f_Q$ of $10^{10-11.5}M_\odot$ satellites  that are quenched, and the morphological mix of these quenched satellites $f_{Q_{ETG}}$, depend on group halo mass, on normalized halo-centric distance and on the surrounding LSS density. 

Similarly to other studies, we find that the total quenched fraction $f_Q$ does not depend much on halo mass or on LSS density.  At the galaxy mass scales that we probe, an average of about 50$\%$ (at low stellar masses with a median $M_{galaxy}\sim10^{10.35} M_\odot$) and $\sim70\%$ (at high stellar masses with a median $M_{galaxy}\sim10^{10.85} M_\odot$) of satellites have been quenched in one way or the other.   The fraction of quenched satellites increases with decreasing distance from the group centres, from $\lta40\%$ at $\sim1.5$ virial radii, up to $\gta60\%$ in the group centres for satellites in the low galaxy  mass bin, and  $\sim50\%$ and $\gta80\%$ for satellites in the high galaxy mass bin.  These results are broadly consistent with previous findings, e.g. van den Bosch et al.\ (2008), Peng et al.\ (2012), Wetzel et al.\ (2012).

The observed trends in $f_Q$ are consistent with a picture of two independent quenching channels, one that  acts on all galaxies and is independent of the environment of a galaxy (i.e., the mass-quenching channel of Peng et al.\ 2010), and another one that acts only on satellites and depends on halo-centric radius but not on halo mass  (the environment- or satellite-quenching suggested by the same, and other, authors).  

The morphological mix in the quenched satellite fraction, quantified by the fraction $f_{Q_{ETG}}$ of quenched satellites that have an early-type morphology, remains on the other hand strikingly constant across all environments, i.e., we detect no significant variations in $f_{Q_{ETG}}$ with halo mass in the range $\sim10^{12.2\rightarrow14.9} M_\odot$, with halo-centric distance out to and beyond the virial radii of the group halos, or with LSS density (at least  within the range probed by ZENS group with masses $<10^{13.7} M_\odot$). At low and high galaxy stellar masses, respectively  about 50$\%$ and 70\% of  quenched satellites have an early-type morphology in halos of all masses and at any halo-centric radius.  The lack of any dependence on the halo-centric radius, despite the clear change in $f_Q$, shows that the morphological mix of quenched galaxies at a given mass is decoupled from the overall quenched fraction.

The lack of any environmental trend with radius of the morphological mix of the quenched satellite population suggests that both mass- and satellite-quenching processes have the same transformational effect on the morphologies of the galaxies and/or that they have no effect at all.
We furthermore note here that the idea of two {\it independent} quenching processes is argued in the 2010 Peng et al.\  analysis on the basis of the separability of their red fraction with mass and environment. It is however possible that the underlying physical mechanisms at work in satellite-quenching could be the same as for mass-quenching, i.e. that they are two different manifestations of the same underlying physical process. One way of doing this would be if satellite quenching is the result of a process linked to the parent halo mass.  The striking mass-independence of satellite quenching could then arise if the distribution of host halo masses  for satellites (at the halo masses of interest) was largely independent of the stellar mass of the satellite, which would be supported by the independence of the satellite mass function on group halo mass (e.g., Peng et al.\ 2012).

Although there appears to be no difference in the morphological outcomes of the two quenching processes, the morphologies of the quenched satellites are not surprisingly systematically different from their star-forming counterparts.  The quenched galaxies have higher $B/T$ and smaller half-light radii.  The morphology-density relation of Dressler (1980) therefore appears to reflect changes in the quenched fraction with radius in the halo rather than changes in the morphological mix of quenched galaxies.

The total $I$-band luminosities and light profiles of the bulges of quenched satellites are very similar to those of the star-forming ones.   Any mass growth in the bulges associated with quenching cannot significantly change their luminosities.  Instead, the changes in $B/T$ and in overall half-light radii are associated with differences in the disks. 

Surface brightness fading of previously star-forming disks after star-formation ceases increases the morphological (light-defined) $B/T$ and decreases the half-light radii, even if there are no underlying structural changes in the stellar mass distributions.  We show that such fading can quantitatively explain the morphological differences in the $B/T$ between star-forming and quenched satellites, and also will reduce the half-light radii.  The quenched satellites also exhibit systematically smaller disk scale lengths, which further shrinks the half-light radii relative to those given by uniform disk-fading. These smaller disk scale lengths could be the result either of differential fading with radius in the disks, e.g. due to age gradients, or if the disks of galaxies were systematically smaller, at a given mass, at earlier times.  Both of these would be expected if individual disks grow inside-out.

The overall conclusion is that, at least at the relatively low redshifts being probed here, neither mass-quenching nor satellite-quenching are likely to be changing the mass-defined B/T, which is therefore probably set by other processes operating prior to the onset of quenching.    This emerging picture from the low redshift analysis is consistent with a merger origin for the most massive, pressure-supported spheroids in the local Universe, and disk-fading of bulge-dominated structures as  a key process leading to the typical rotation-supported S0+Elliptical population around $M*$. Of course, the situation at much higher redshifts may or may not be the same.

Current SAMs postulate that early-type morphologies are the result of mergers preceding the infall of the satellite into the virialized group potential, and that subsequent  group halo physics, in the form of gradual removal of galaxies' hot gas halos, leaves the galaxies morphologies unchanged. This picture is not without challenges, not only because the models overpredict the total quenched satellite fraction in all environments, but also because the role of mergers and the quantitative merger rates need to be better defined as a function of cosmic epoch. Specifically, merger rates need to be  made consistent with the emerging paradigm of a non-merging, star-forming galaxy main sequence (e.g., Daddi et al 2007) whose star-formation is fed in quasi-steady-state, from early times, by direct accretion of relatively cold gas streams (e.g., Dekel et al.\ 2009; see also Dave' et al.\ 2012 and Lilly et al.\ 2013). Nevertheless,  SAMs which include a  progressive  removal of the hot gas reservoir of galaxies within group halos well reproduce both the lack of dependence on halo mass and   the dependence on halo-centric distance of the quenched satellite fraction that we see in our ZENS data.  In all environments, they over-predict however  the fraction of quenched satellites with a dominant disk component. Overall, these model successes and failures in  matching the observed quenched satellite  population at redshift zero indicate  that  further model improvements should involve not only the  physical recipes for gas removal from galaxies in heavy halos, but also their star formation and mass assembly histories as a function of galaxy mass.

\acknowledgments
CMC thanks Alvio Renzini for comments and discussions, and Peter Capak, Nick Scoville, and the Caltech Astronomy Institute for the warm hospitality and the enjoyable and stimulating atmosphere during their sabbatical visit, when this work has been finalised. We gratefully acknowledge support by the Swiss National Science Foundation. 
This publication makes use of data from ESO Large Program 177.A-0680. 
We also use data products from the Two Micron All Sky Survey, 
which is a joint project of the University of Massachusetts and the Infrared Processing 
and Analysis Center/California Institute of Technology, funded by the National Aeronautics and
Space Administration and the National Science Foundation. 
This research has made use of the NASA/IPAC Extragalactic Database (NED) which is operated by the Jet Propulsion Laboratory, California Institute of Technology, under contract with the National Aeronautics and Space Administration.

\clearpage

\LongTables
\begin{sidewaystable}
\begin{deluxetable}{c |cccc | ccc}
\tabletypesize{\small}
\tablewidth{0pt}\setcounter{table}{1}
\tablecaption{\label{t1}Fraction $f_Q$ of quenched satellites, and $f_{Q_{ETG}}$ of quenched satellites with an early-type morphology, in the  $M_{Halo}$-$R/R_{vir}$ bins of Figures \ref{fA1} and \ref{fA2}}
\tablehead{
\multicolumn{1}{c|}{} & \multicolumn{4}{c|}{$10^{10}\leqslant \log_{10}(M/\Msol)<10^{10.7}$}  & \multicolumn{3}{c}{$10^{10.7}\leqslant \log_{10}(M/\Msol)\leqslant10^{11.5}$} 
}
\startdata
& & & & & & & \\

 $ \log_{10}(M/\Msol)\rightarrow$ & All $M_{Halo}$ & $ [12.2,13.2[ $ & $[13.2,13.7[$  & $[13.7,14.8]$   &  All $M_{Halo}$ & $[12.8,13.7[$  & $[13.7,14.8]$ \\
  &&&&&&&\\
 \cline{1-8}
\cline{1-8}
 \\
 \% quenched  [$f_Q$]  \\
  \\
\cline{1-8}\\
All $R$            && \grey{0.58$^{+0.05}_{-0.08}$(0.56)}    &    \grey{0.56$^{+0.05}_{-0.06}$(0.57)}   & \grey{0.54$^{+0.04}_{-0.03}$(0.59)}   && \grey{0.69$^{+0.08}_{-0.12}$(0.71)}   &  \grey{0.77$^{+0.04}_{-0.07}$(0.80)} \\ 
                       &                                       & {0.54$^{+0.04}_{-0.06}$}  &    {0.50$^{+0.04}_{-0.04}$}   & {0.51$^{+0.04}_{-0.03}$}  && {0.73$^{+0.07}_{-0.08}$} & {0.75$^{+0.04}_{-0.07}$} \\
$R\leqslant 0.5R_{vir}$    & \grey{0.64$^{+0.04}_{-0.04}$(0.67)}  & \grey{0.55$^{+0.09}_{-0.11}$(0.53)}                                               & \grey{0.84$^{+0.04}_{-0.11}$(0.87)}    & \grey{0.62$^{+0.06}_{-0.05}$(0.67)}       &  \grey{0.88$^{+0.03}_{-0.06}$(0.87)}& \grey{0.75$^{+0.07}_{-0.18}$(0.72)}   & \grey{0.93$^{+0.03}_{-0.07}$(0.92)} \\ 
                       &    {0.61$^{+0.03}_{-0.04}$}       & {0.56$^{+0.08}_{-0.10}$}   &   {0.68$^{+0.06}_{-0.10}$}    & {0.60$^{+0.05}_{-0.05}$} &  {0.86$^{+0.03}_{-0.06}$}& {0.77$^{+0.07}_{-0.17}$} &{0.90$^{+0.03}_{-0.07}$} \\
$0.5R_{vir}<R\leqslant R_{vir}$ &  \grey{0.52$^{+0.04}_{-0.04}$(0.52)}                                 & \grey{0.61$^{+0.08}_{-0.12}$(0.59)}    & \grey{0.42$^{+0.09}_{-0.08}$(0.41)}      & \grey{0.54$^{+0.06}_{-0.06}$(0.56)}   &  \grey{0.71$^{+0.06}_{-0.09}$(0.78)}& \grey{0.65$^{+0.14}_{-0.20}$(0.80)}  & \grey{0.73$^{+0.06}_{-0.11}$(0.77)} \\
                         &    {0.48$^{+0.04}_{-0.04}$}&   {0.55$^{+0.08}_{-0.11}$} & {0.39$^{+0.08}_{-0.07}$}     & {0.50$^{+0.06}_{-0.05}$}   &  {0.76$^{+0.05}_{-0.08}$}& {0.77$^{+0.09}_{-0.17}$} & {0.76$^{+0.05}_{-0.11}$} \\
$R> R_{vir}$            &   \grey{0.44$^{+0.05}_{-0.05}$(0.49)}                                                         & \grey{0.59$^{+0.10}_{-0.15}$(0.59)}   & \grey{0.47$^{+0.09}_{-0.10}$(0.48)}       & \grey{0.38$^{+0.08}_{-0.08}$(0.45)}   &  \grey{0.21$^{+0.17}_{-0.08}$(0.28)}& \grey{0.46$^{+0.29}_{-0.21}$(0.46)}  & \grey{0.14$^{+0.21}_{-0.05}$(0.20)} \\ 
                         &  {0.45$^{+0.04}_{-0.04}$} &  {0.53$^{+0.07}_{-0.08}$}  & {0.46$^{+0.07}_{-0.06}$}     & {0.36$^{+0.07}_{-0.07}$}   &  {0.47$^{+0.10}_{-0.09}$}& {0.66$^{+0.14}_{-0.13}$} & {0.33$^{+0.13}_{-0.11}$} \\
 && & & & & &\\
\cline{1-8}\\
 \% quenched ETG [$f_{Q_{ETG}}$]   \\
 \\
\cline{1-8}\\
   & & &  & && & \\
All $R$          &                                          & \grey{0.50$^{+0.08}_{-0.09}$(0.47)}   & \grey{0.49$^{+0.08}_{-0.07}$(0.45)}   & \grey{0.52$^{+0.05}_{-0.05}$(0.52)}  && \grey{0.68$^{+0.10}_{-0.14}$(0.66)}  & \grey{0.72$^{+0.06}_{-0.07}$(0.70)} \\
                                &                                  & {0.54$^{+0.06}_{-0.08}$}   & {0.54$^{+0.06}_{-0.06}$} & {0.51$^{+0.05}_{-0.04}$} && {0.78$^{+0.07}_{-0.10}$} & {0.69$^{+0.06}_{-0.06}$} \\
$R\leqslant 0.5R_{vir}$       &   \grey{0.51$^{+0.05}_{-0.05}$(0.49)}               & \grey{0.47$^{+0.12}_{-0.13}$(0.43)}     & \grey{0.49$^{+0.12}_{-0.09}$(0.46)} & \grey{0.53$^{+0.07}_{-0.06}$(0.52)}  &  \grey{0.71$^{+0.06}_{-0.08}$(0.67)} & \grey{0.61$^{+0.15}_{-0.16}$(0.55)}  & \grey{0.74$^{+0.08}_{-0.08}$(0.71)} \\
                                     &{0.51$^{+0.05}_{-0.05}$}     &   {0.51$^{+0.11}_{-0.13}$} & {0.51$^{+0.11}_{-0.09}$} & {0.51$^{+0.07}_{-0.06}$} &  {0.68$^{+0.06}_{-0.07}$} & {0.65$^{+0.13}_{-0.16}$} & {0.69$^{+0.08}_{-0.09}$} \\
$0.5R_{vir}<R\leqslant R_{vir}$    &  \grey{0.52$^{+0.06}_{-0.06}$(0.50)}  & \grey{0.47$^{+0.12}_{-0.14}$(0.43)}    & \grey{0.39$^{+0.14}_{-0.11}$(0.33)}  & \grey{0.57$^{+0.08}_{-0.08}$(0.63)}  &  \grey{0.68$^{+0.07}_{-0.10}$(0.70)} & \grey{0.76$^{+0.09}_{-0.29}$(0.76)}  & \grey{0.67$^{+0.09}_{-0.12}$(0.68)} \\ 
                                  &    {0.55$^{+0.05}_{-0.05}$}                              &  {0.51$^{+0.11}_{-0.14}$}  & {0.47$^{+0.12}_{-0.11}$} & {0.59$^{+0.07}_{-0.07}$} & {0.68$^{+0.07}_{-0.09}$} & {0.73$^{+0.09}_{-0.22}$}  & {0.67$^{+0.08}_{-0.12}$} \\
$R> R_{vir}$                      &      \grey{0.48$^{+0.08}_{-0.08}$(0.48)}                & \grey{0.61$^{+0.15}_{-0.16}$(0.61)}   & \grey{0.59$^{+0.11}_{-0.15}$(0.55)}  & \grey{0.31$^{+0.15}_{-0.09}$(0.33)}  &  \grey{1.00$_{-0.44}$(1.00)} & \grey{1.00$_{-0.60}$(1.00)}   & \grey{1.00$_{-0.60}$(1.00)} \\ 
                                   &   {0.52$^{+0.06}_{-0.06}$}    &  {0.59$^{+0.09}_{-0.11}$} & {0.61$^{+0.08}_{-0.10}$} & {0.30$^{+0.13}_{-0.09}$} &  {0.90$^{+0.03}_{-0.15}$} & {1.00$_{-0.21}$}  & {0.75$^{0.11}_{-0.27}$} \\

\enddata
\tablecomments{\\ The table shows the fraction $f_Q$ of quenched satellites of any morphology (top), and the fraction $f_{Q_{ETG}}$ of quenched galaxies which have an early-type morphology (bottom),  in each of the quadrants in which Figures  \ref{fA1} and \ref{fA2} split  the  $M_{Halo}$-$R/R_{vir}$ plane,  for  the two galaxy stellar mass bins of our analysis. Black entries show values obtained considering the entire ZENS sample of relaxed plus unrelaxed groups; grey entries refer to fractions derived when using only the sample of  relaxed groups.  Always in grey, in parenthesis, are reported the corresponding fractions for the `clean' sample of relaxed groups only: this excludes groups for which a self-consistent solution for the central galaxy in the mass, halo-centric distance and velocity parameter space  could be found within the errors on the galaxy mass estimates, but the galaxy that satisfies these conditions  is not the galaxy with the nominally highest `best-fit' stellar mass (see Section \ref{survey} and Paper I).
All fractions are corrected for the effects of the 2dFGRS spectroscopic incompleteness; we highlight that the applied corrections are quite small. The errors  indicate bayesian $\pm 1 \sigma$ confidence intervals for a binomial distribution.
 }
\end{deluxetable}
\end{sidewaystable}
\clearpage

\clearpage 
\begin{deluxetable*}{c |ccc | cc}
\tabletypesize{\small}
\tablewidth{0pt}
\tablecaption{Fraction $f_Q$ of quenched satellites,  and $f_{Q_{ETG}}$ of quenched satellites with an early-type morphology,  as a function of  $\delta_{LSS}$   (in the low galaxy mass bin also split in the   $R/R_{vir}$ bins of Figures  \ref{fA3} and \ref{fA4})}
\tablehead{
\multicolumn{1}{c|}{}   & \multicolumn{3}{c|}{$10^{10}\leqslant \log_{10}(M/\Msol)<10^{10.7}$}  & \multicolumn{2}{c}{$10^{10.7}\leqslant \log_{10}(M/\Msol)\leqslant10^{11.5}$}
}
\startdata
   & & & & \\
$\log_{10}(1+\delta_{LSS})\rightarrow$ & [-0.5,0.3[ & [0.3,0.7[ & [0.7,2] & [-0.5,0.7[ &  [0.7,2]\\
   & & & & \\
\cline{1-6}
 \\
  \% quenched [$f_Q$]   \\
   \\ 
   \cline {1-6} \\
All $R$     &     \grey{0.48$^{+0.08}_{-0.08}$(0.48)}     & \grey{0.64$^{+0.06}_{-0.07}$(0.61)} & \grey{0.57$^{+0.05}_{-0.06}$(0.59)}  & \grey{0.60$^{+0.12}_{-0.15}$(0.60)}  & \grey{0.77$^{+0.08}_{-0.18}$(0.85)}\\
      & {0.47$^{+0.06}_{-0.06}$}       & {0.53$^{+0.05}_{-0.05}$} & {0.52$^{+0.05}_{-0.05}$} & {0.63$^{+0.09}_{-0.12}$} & {0.85$^{+0.05}_{-0.13}$} \\
$R\leqslant 0.5R_{vir}$     &     \grey{0.54$^{+0.10}_{-0.14}$(0.54)}     & \grey{0.77$^{+0.08}_{-0.13}$(0.76)} & \grey{0.75$^{+0.06}_{-0.13}$(0.77)}  & & \\
      & {0.56$^{+0.09}_{-0.13}$}       & {0.66$^{+0.09}_{-0.11}$} & {0.64$^{+0.07}_{-0.12}$} & &\\
$0.5R_{vir}<R\leqslant R_{vir}$     &     \grey{0.36$^{+0.16}_{-0.11}$(0.36)}     & \grey{0.54$^{+0.11}_{-0.14}$(0.51)} & \grey{0.52$^{+0.08}_{-0.09}$(0.52)}  & & \\
      & {0.39$^{+0.14}_{-0.11}$}       & {0.43$^{+0.11}_{-0.10}$} & {0.49$^{+0.08}_{-0.09}$} & &\\
$R> R_{vir}$    &     \grey{0.52$^{+0.14}_{-0.18}$(0.52)}     & \grey{0.56$^{+0.12}_{-0.12}$(0.56)} & \grey{0.46$^{+0.10}_{-0.12}$(0.48)}  & & \\
      & {0.44$^{+0.11}_{-0.10}$}       & {0.51$^{+0.07}_{-0.06}$} & {0.48$^{+0.07}_{-0.09}$} & &\\
   & & & & \\
\cline{1-6}
\\
\% quenched ETG [$f_{Q_{ETG}}$] \\
   \\
\cline{1-6}
   & & & & \\
All $R$        &     \grey{0.45$^{+0.11}_{-0.10}$(0.45)} & \grey{0.52$^{+0.09}_{-0.09}$(0.48)} & \grey{0.49$^{+0.07}_{-0.07}$(0.45)} & \grey{0.84$^{+0.06}_{-0.22}$(0.84)}  & \grey{0.56$^{+0.14}_{-0.17}$(0.48)}\\
     & {0.51$^{+0.09}_{-0.09}$}   & {0.59$^{+0.06}_{-0.07}$} & {0.51$^{+0.06}_{-0.07}$} & {0.91$^{+0.03}_{-0.15}$} & {0.66$^{+0.10}_{-0.14}$} \\
$R\leqslant 0.5R_{vir}$     &     \grey{0.36$^{+0.19}_{-0.12}$(0.36)}     & \grey{0.54$^{+0.13}_{-0.14}$(0.50)} & \grey{0.50$^{+0.12}_{-0.12}$(0.47)}  & & \\
      & {0.39$^{+0.17}_{-0.11}$}       & {0.57$^{+0.12}_{-0.13}$} & {0.53$^{+0.11}_{-0.13}$} & &\\
$0.5R_{vir}<R\leqslant R_{vir}$     &     \grey{0.48$^{+0.22}_{-0.19}$(0.48)}     & \grey{0.43$^{+0.18}_{-0.14}$(0.33)} & \grey{0.43$^{+0.11}_{-0.12}$(0.37)}  & & \\
      & {0.59$^{+0.16}_{-0.21}$}       & {0.44$^{+0.17}_{-0.13}$} & {0.49$^{+0.10}_{-0.13}$} & &\\
$R> R_{vir}$    &     \grey{0.70$^{+0.15}_{-0.23}$(0.70)}     & \grey{0.51$^{+0.15}_{-0.17}$(0.51)} & \grey{0.62$^{+0.13}_{-0.18}$(0.57)}  & & \\
      & {0.65$^{+0.12}_{-0.16}$}       & {0.64$^{+0.07}_{-0.11}$} & {0.51$^{+0.11}_{-0.13}$} & &\\
\enddata
\tablecomments{The table shows the fraction $f_Q$ of quenched satellites,  and the fraction $f_{Q_{ETG}}$ of quenched satellites with an early-type morphology as a function of $\delta_{LSS}$ (in the low galaxy mass bin, also split  in   bins of  $R/R_{vir}$ as in  Figures  \ref{fA4} and \ref{fA4}).    Note that in this analysis only groups with $M_{Halo}\leq10^{13.7}M_\odot$ are considered. All entries are  coded as in  Table \ref{t1}. }\label{t2}
\end{deluxetable*}

\appendix

\section{Quenched satellite  fractions and the morphologies of quenched satellites in relaxed and unrelaxed groups}\label{allrelaxed}

In ZENS about 40\% of groups in the sample are defined as unrelaxed. As discussed in Section \ref{survey}, this means it is not possible  to identify a single galaxy with a stellar mass, location within the group and velocity relative to the mean which can self-consistently be ranked as the central galaxy in the group.  In Paper I we argued that  a substantial fraction of these unrelaxed groups can be the result of contamination from interloper galaxies, although about a quarter to a third of them are most likely genuinely young/merging groups.   

\begin{figure*} 
\begin{center}
\includegraphics[width=0.45\textwidth]{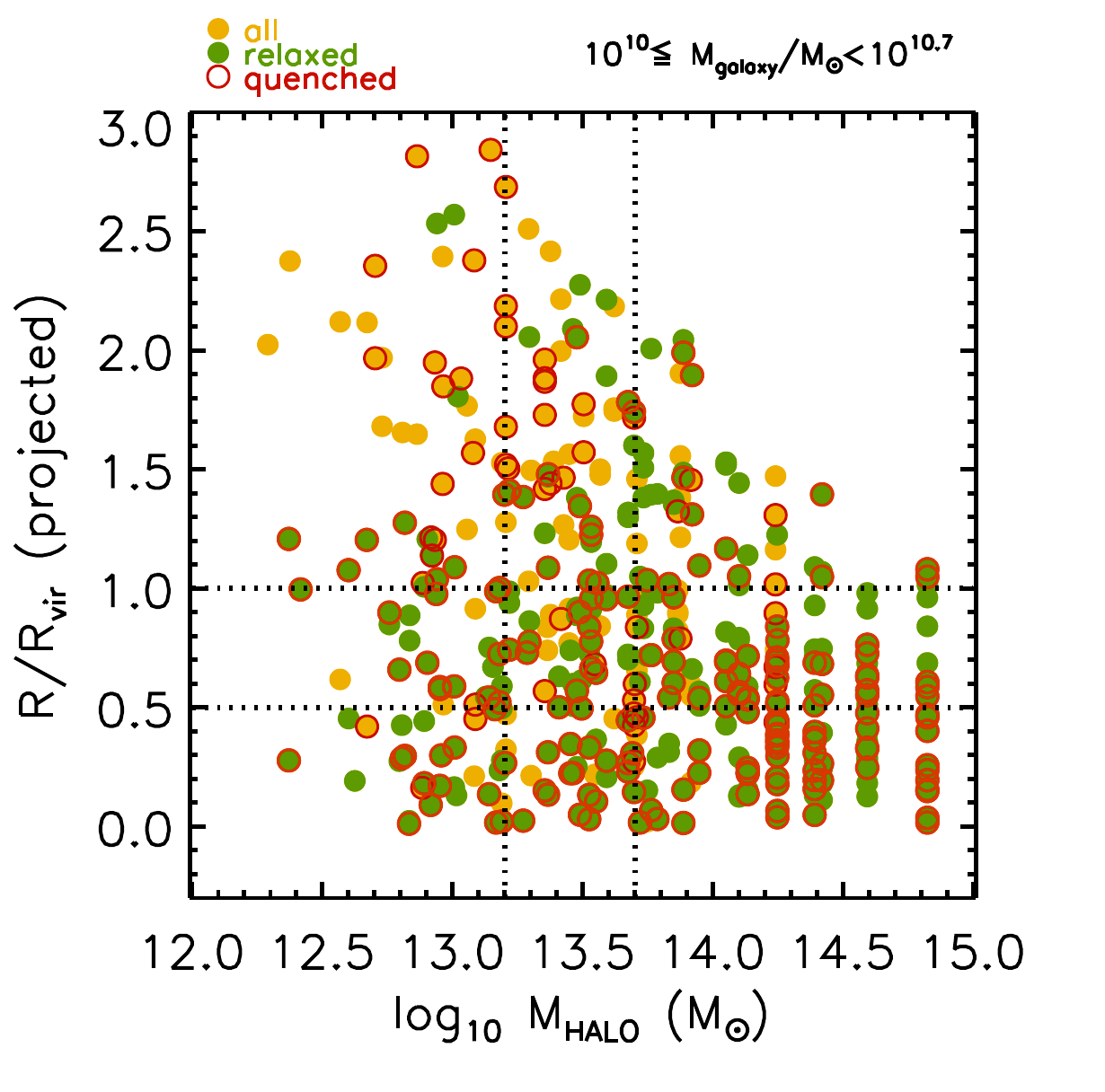}
\includegraphics[width=0.45\textwidth]{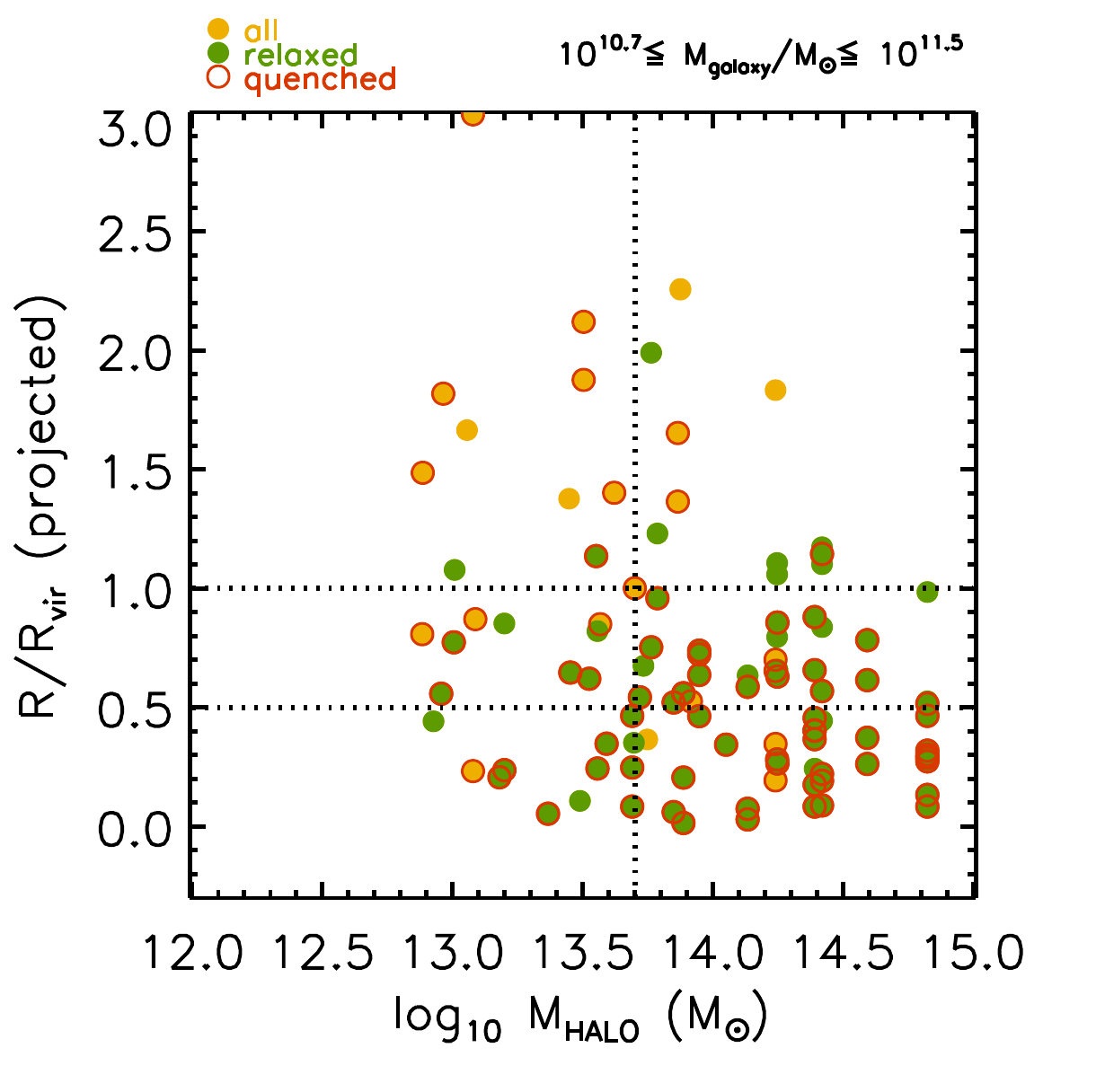}
\caption[]{{\it Left:} Satellite galaxies with stellar mass in the range $10^{10-10.7}M_{\odot}$ are plotted as dots in the halo-centric distance versus halo mass   plane. Yellow dots are satellites in ZENS groups  which are classified as unrelaxed (see text in Appendix \ref{allrelaxed} and Paper I) and green dots are those in dynamically-relaxed groups. Surrounding red rings identify the {\it quenched} satellites, independent of their morphologies. {\it Right:} As in left panel, but for satellites in the higher stellar mass bin $10^{10.7}-10^{11.5} M_{\odot}$. \label{fA1}}
\end{center}
\end{figure*}

To establish  whether the groups defined as unrelaxed (regardless of the reason) have an impact on the fractions of   quenched satellites in the different environments,  we plot  in Figure  \ref{fA1}  the $R/R_{vir}$ versus $M_{Halo}$ plane for the two galaxy mass bins of our analysis. Each dot in these plots is a satellite galaxy: yellow dots are satellites in unrelaxed groups, and green dots indicate satellites in relaxed groups. Red circles around yellow and green symbols indicate that the given satellite is a quenched galaxy. The top half of Table \ref{t1} shows the values for the fractions $f_Q$ of satellites that are quenched  in each of the bins in the $M_{Halo}$ vs. $R/R_{vir}$ planes of Figure   \ref{fA1} that are delineated by the dotted lines. The table separately lists, for  each galaxy stellar mass bin, the fractions $f_Q$ that are obtained when using the different group samples. 
The errors in Table \ref{t1} indicate bayesian $\pm 1 \sigma$ confidence intervals for a binomial distribution
(as e.g., Cameron  2011  and references therein).  
All fractions have been corrected for spectroscopic incompleteness in the parent 2dFGRS catalog. These corrections are small, and are done by calculating the effective number of galaxies of a given category as the sum of the inverse galaxy weights $w_i$ defined in Appendix A of Paper I, and characterising the 2dFGRS completeness at the galaxy positions. 
For example, the effective  number of quenched galaxies is equal to $N_{Q}=\sum_i 1/w_{Q,i}$.

\begin{figure*} 
\begin{center}
\includegraphics[width=0.45\textwidth]{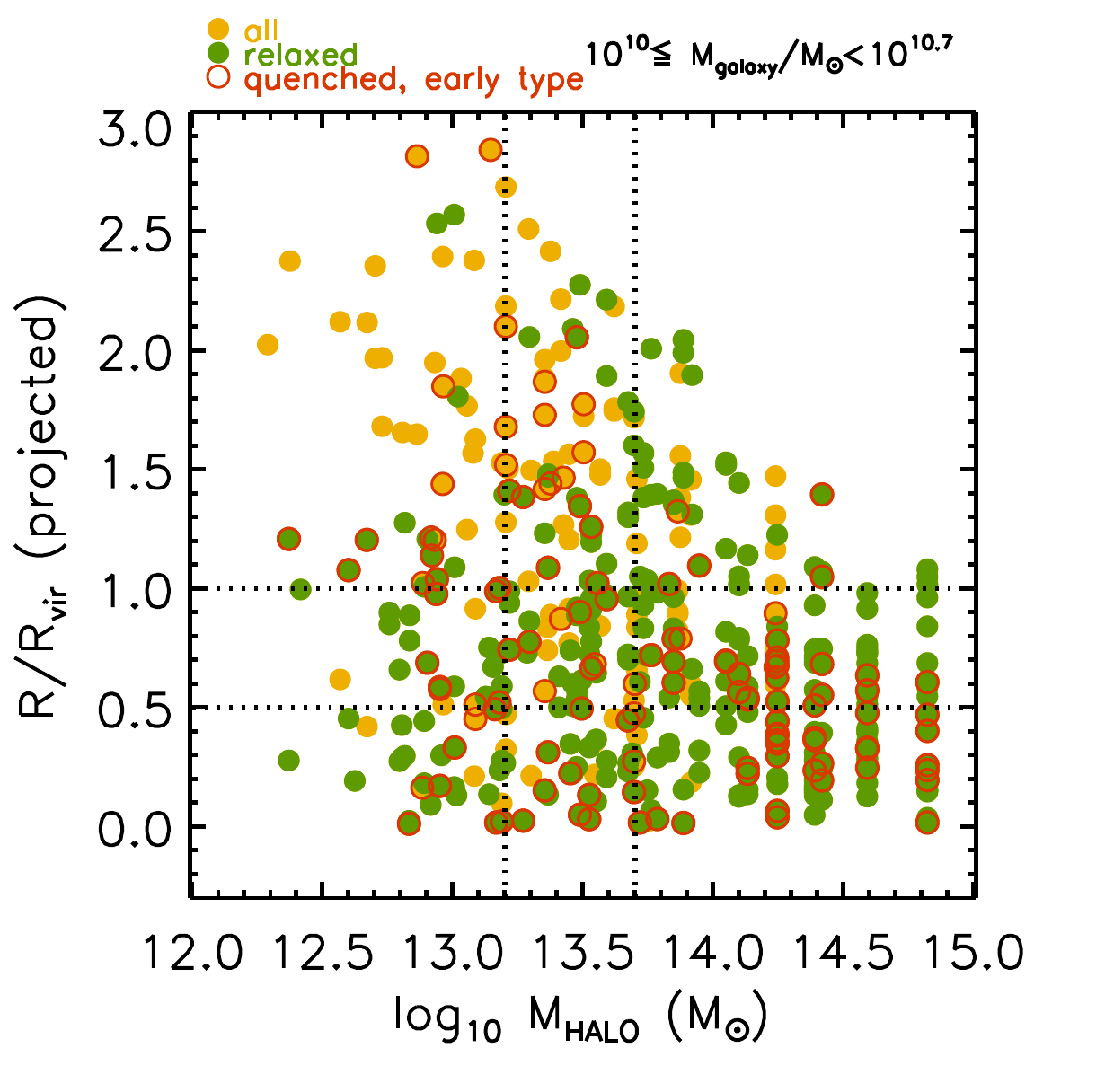}
\includegraphics[width=0.45\textwidth]{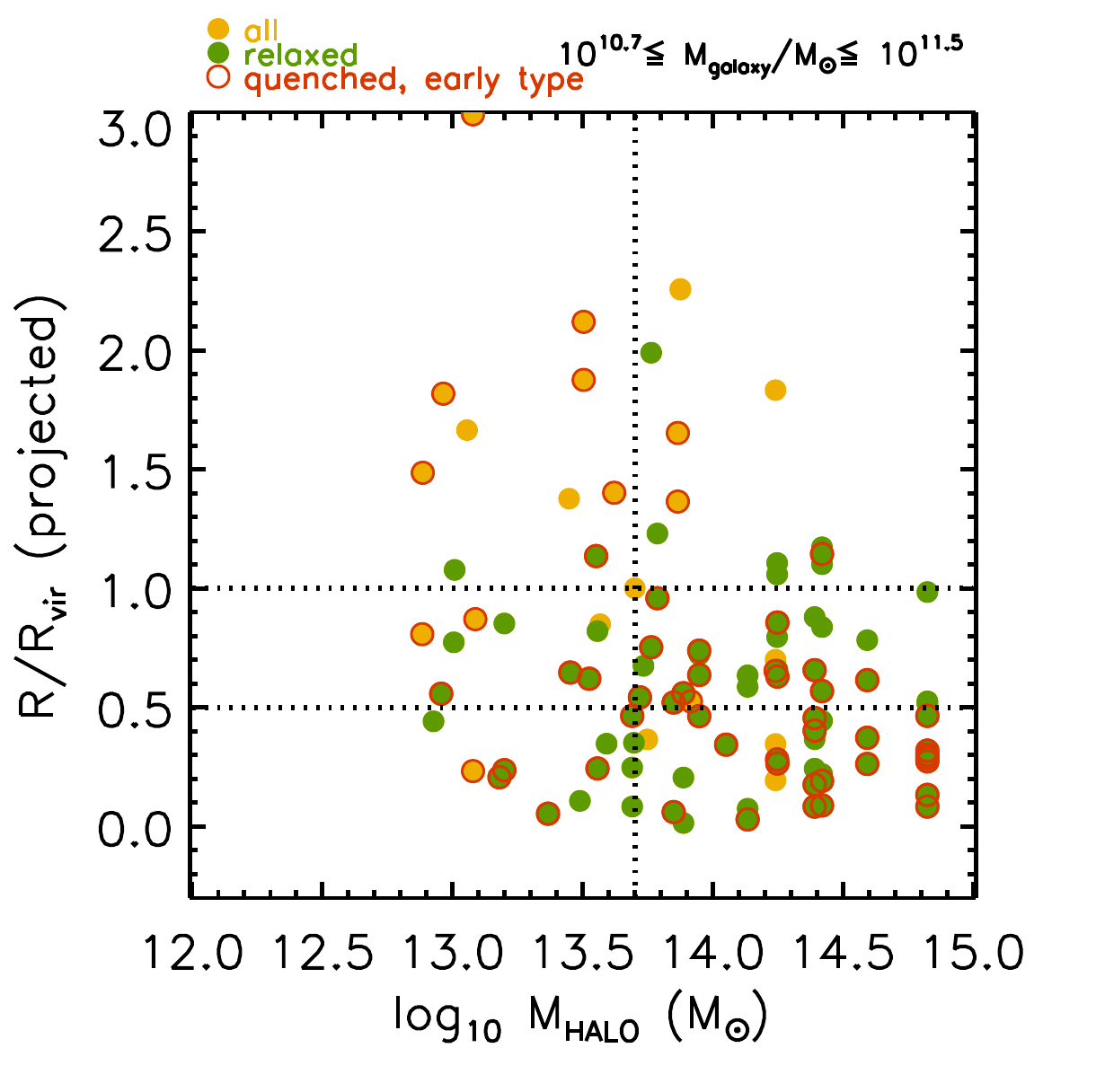}
\caption[]{As in Figure \ref{fA1}, but now the red rings identify quenched satellites that have an early-type morphology.}\label{fA2}
\end{center}
\end{figure*}

\begin{figure*} 
\begin{center}
\includegraphics[width=0.45\textwidth]{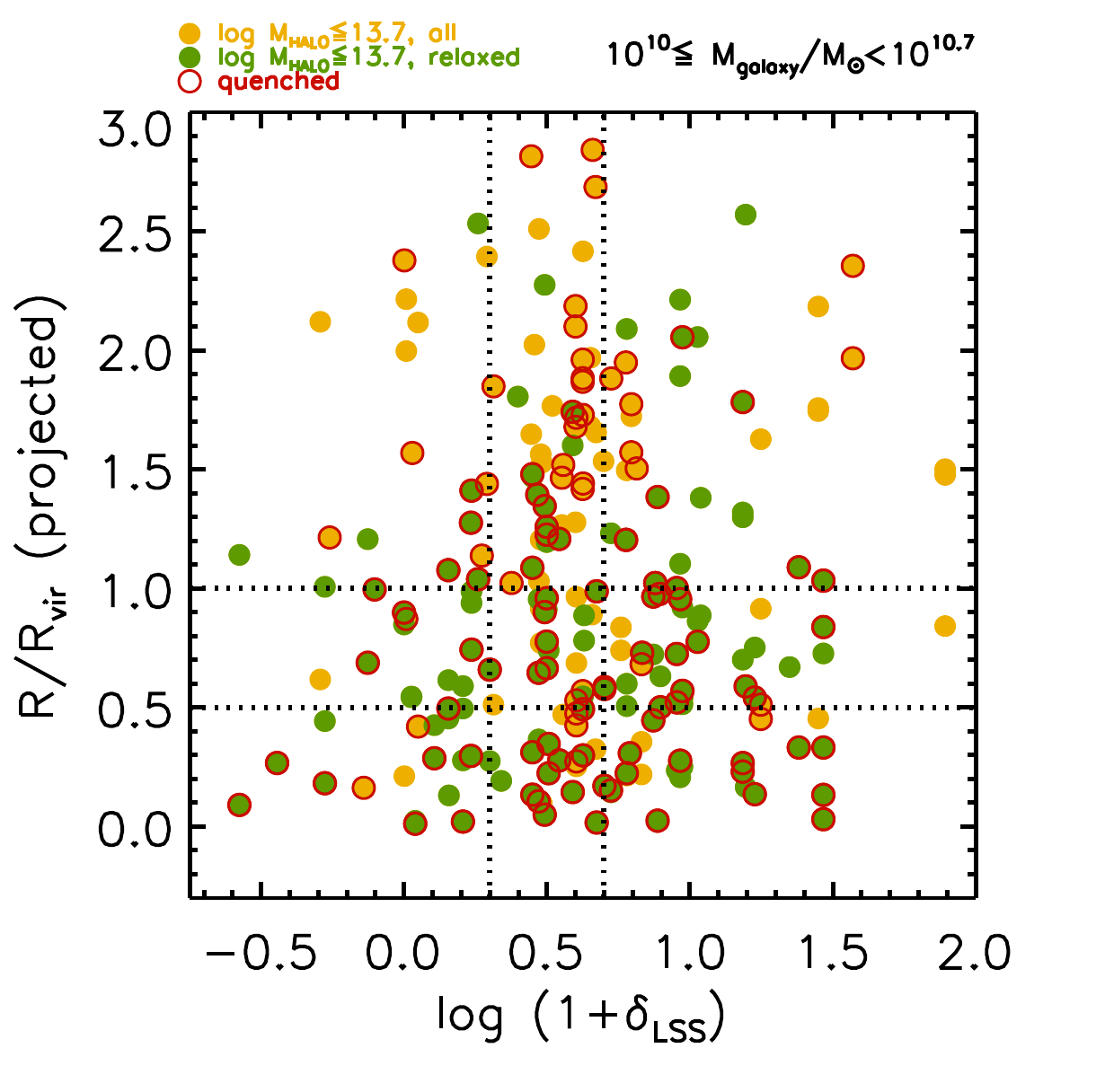}
\includegraphics[width=0.45\textwidth]{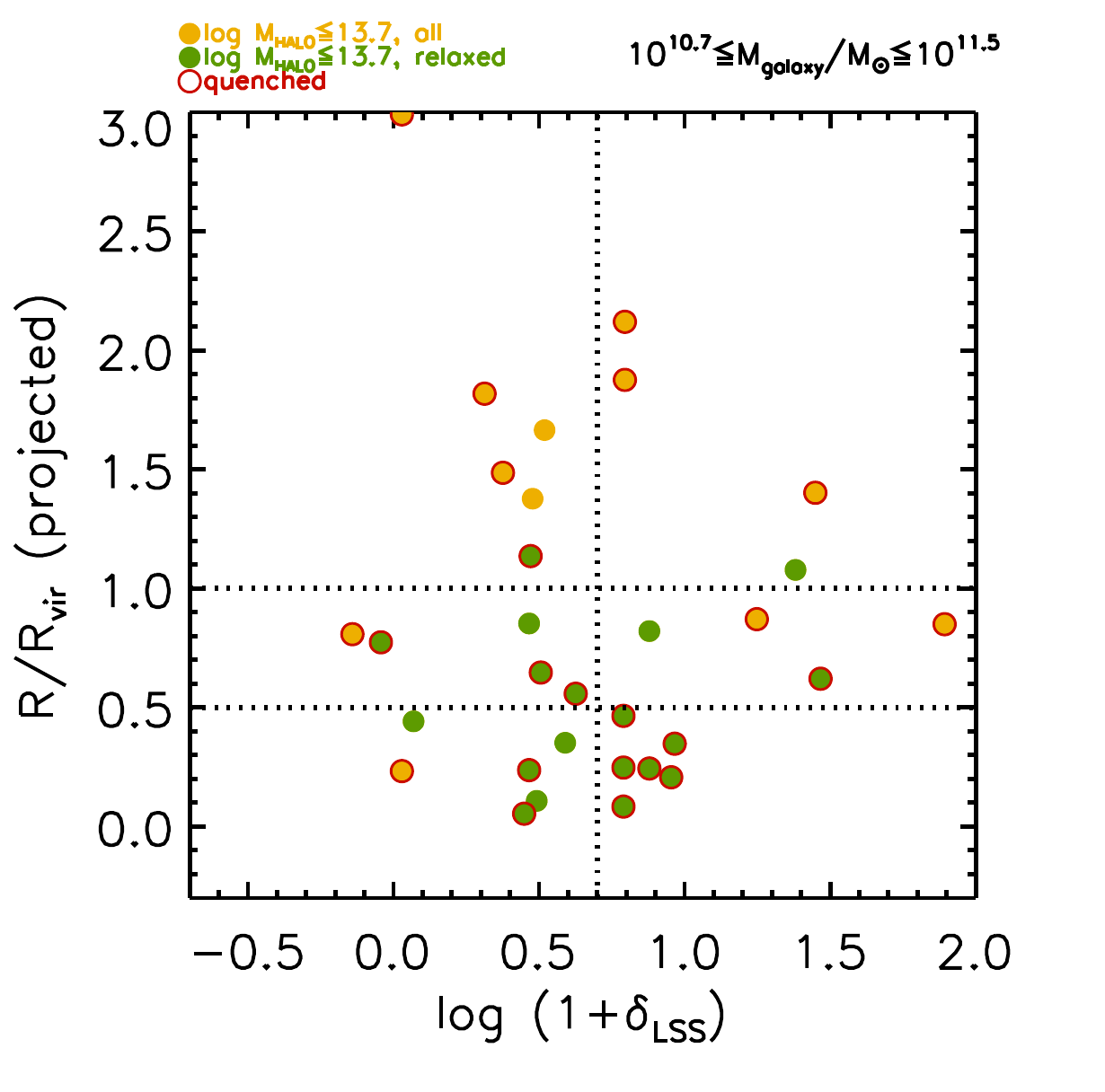}
\caption[]{As in Figure \ref{fA2}, but now using the LSS over-density as the environment parameter on the horizontal axis. Red rings identify quenched satellites, independent of their morphology.}\label{fA3}
\end{center}
\end{figure*}

\begin{figure*} 
\begin{center}
\includegraphics[width=0.45\textwidth]{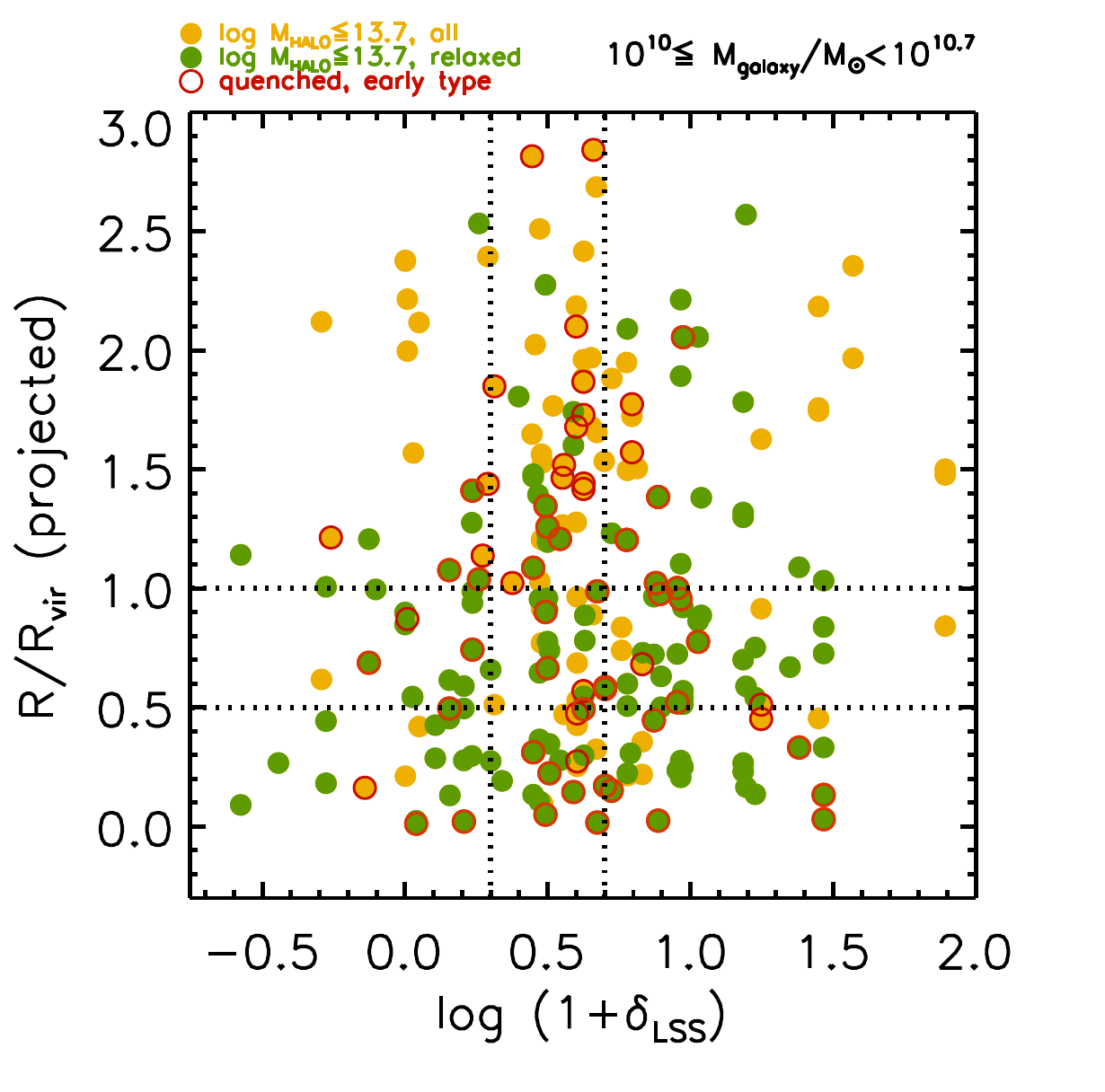}
\includegraphics[width=0.45\textwidth]{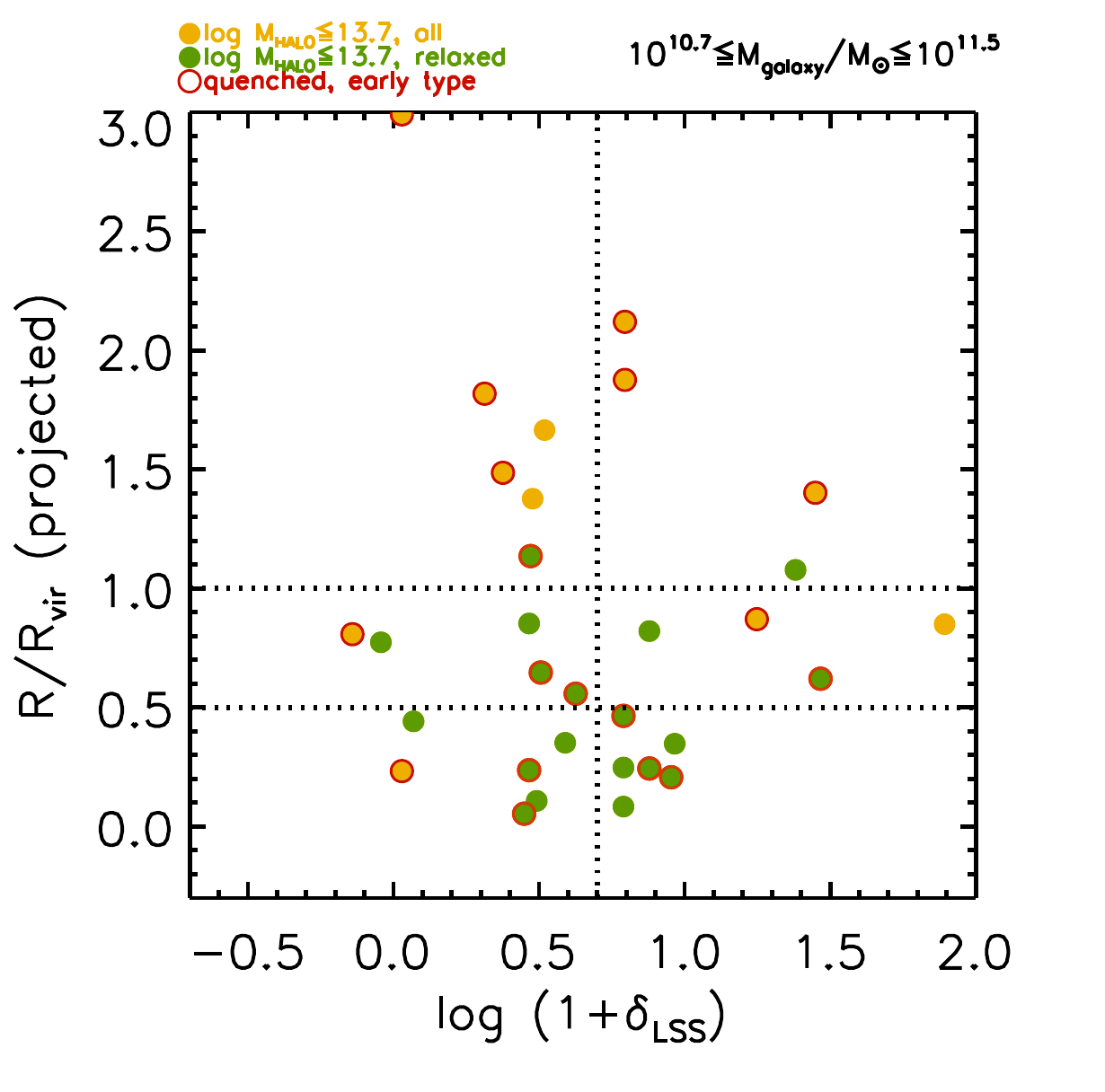}
\caption[]{As in Fig \ref{fA3}, but with the red rings now indicating quenched galaxies with early type morphology.} \label{fA4}
\end{center}
\end{figure*}

Figure  \ref{fA1} is repeated as Figure  \ref{fA2}, but now the red circles indicate the quenched satellites that have an {\it early-type morphology}. Correspondingly, the bottom part of Table \ref{t1} reports the fractions $f_{Q_{ETG}}$ of quenched satellites with an early-type morphology, relative to the total population of quenched satellites in each given $R/R_{vir}$-$M_{Halo}$  bin, again for the whole ZENS sample and for the sample of relaxed  groups (and `clean relaxed' groups, see Section \ref{survey}).  Figures  \ref{fA3} and \ref{fA4} show satellites in unrelaxed (yellow dots) and relaxed (green dots) groups, in different locations of the $R/R_{vir}$ versus $\delta_{LSS}$ plane, limiting the group sample to halo masses $M_{Halo} \leq 10^{13.7} M_\odot$ (for the reason discussed in Section \ref{survey}). Correspondingly, for the low galaxy stellar mass bin, Table \ref{t2}  lists the fractions of quenched satellites, and the fraction of quenched satellites with an early-type morphology (relative to the total), in the  bins of $\delta_{LSS}$ and $R/R_{vir}$ shown in the left panels of Figures \ref{fA3} and \ref{fA4} (integrated over all  $M_{Halo} \leq 10^{13.7} M_\odot$ groups).  At high galaxy stellar masses, the ZENS $M_{Halo} \leq 10^{13.7} M_\odot$  group sample is too small to allow for a simultaneous split of the galaxy sample in both $\delta_{LSS}$ and $R/R_{vir}$. At these high galaxy stellar masses  we therefore only investigate in this paper the total fraction of quenched satellites  and their morphologies in separate $\delta_{LSS}$ bins, by integrating not only over $M_{Halo} \leq 10^{13.7} M_\odot$ groups, but also over all $R/R_{vir}$ values.

The inspection of Figures \ref{fA1}-\ref{fA4} shows that, at  low halo masses and LSS  densities, and in both galaxy stellar mass bins, it is the unrelaxed groups that are the dominant source of galaxies at large $R > R_{vir}$ halo-centric distances. This is not surprising, given that  such unrelaxed systems are either structures in the process of merging, or unreliable physical associations with a misidentified central galaxy (and therefore group center). The nominal $R/R_{vir}$ values for galaxies in these systems are not therefore as meaningful as for the relaxed groups.

We note however that, as shown in Tables \ref{t1} and \ref{t2}, the quenched  fractions in most environmental bins are stable to the inclusion of the unrelaxed groups. The morphological compositions of the quenched satellite populations in the various environmental bins are also, within the errors, very similar, whether or not the unrelaxed groups are included in the analysis. The largest discrepancies are observed at high stellar masses;  however the statistical errors on the estimates make even these discrepancies only marginally significant. 

We therefore conclude that both the quenched fraction of satellites, and the morphological composition of this quenched population of satellites are largely independent of the implied dynamical state of the host group halos.  In the analysis presented in this paper we will thus consider the entire sample of relaxed plus unrelaxed ZENS groups so as to maximise the statistical significance of  our results. 

\end{document}